\documentclass[%
superscriptaddress,
frontmatterverbose, 
bibnotes,
 amsmath,amssymb,
 aps,
floatfix,
]{revtex4-2}

\usepackage{graphicx}
\usepackage{epstopdf, epsfig}
\usepackage[percent]{overpic}

\usepackage{blindtext}
\usepackage{graphicx}
\usepackage{float}
\usepackage{bm}
\usepackage{multirow}
\usepackage{xcolor}

\usepackage[utf8]{inputenc}
\usepackage{xcolor}
\usepackage{float}
\usepackage{subcaption}
\usepackage{amsmath}
\usepackage[left=1.5cm, right=1.5cm, top=1.785cm, bottom=2.0cm]{geometry}
\usepackage{mathptmx}
\usepackage{amsmath,amssymb}
\usepackage{graphicx} 
\usepackage{lastpage}
\usepackage[format=plain,justification=justified,singlelinecheck=false,labelfont=bf,labelsep=space]{caption}
\usepackage{float}
\usepackage[english]{babel}
\begin{document}

\preprint{APS/123-QED}

\title{Rod-climbing rheometry revisited}

\title{Rod-climbing rheometry revisited}
\author{Rishabh V. More}
\affiliation{Department of Mechanical Engineering, Massachusetts Institute of Technology, Cambridge, Massachusetts, USA}
\author{Reid Patterson}
\affiliation{The Lubrizol Corporation, 29400 Lakeland Blvd., Wickliffe, Ohio 44092,USA}

\author{Eugene Pashkovski}
\affiliation{The Lubrizol Corporation, 29400 Lakeland Blvd., Wickliffe, Ohio 44092,USA}

\author{Gareth H. McKinley}
\affiliation{Department of Mechanical Engineering, Massachusetts Institute of Technology, Cambridge, Massachusetts, USA}
\email{gareth@mit.edu}

\date{\today}

\begin{abstract}

The rod-climbing or “Weissenberg” effect in which the free surface of a complex fluid climbs a thin rotating rod is a popular and convincing experiment demonstrating the existence of elasticity in polymeric fluids. The interface shape and steady-state climbing height depend on the rotation rate, fluid elasticity (through the presence of normal stresses), surface tension, and inertia. By solving the equations of motion in the low rotation rate limit for a second-order fluid, a mathematical relationship between the interface deflection and the fluid material functions, specifically the first and second normal stress differences, emerges. This relationship has been used in the past to measure the climbing constant, a combination of the first $(\Psi_{1,0})$ and second $(\Psi_{2,0})$ normal stress difference coefficients from experimental observations of rod-climbing in the low inertia limit. However, a quantitative reconciliation of such observations with the capabilities of modern-day torsional rheometers is lacking. To this end, we combine rod-climbing experiments with both small amplitude oscillatory shear flow measurements and steady shear measurements of the first normal stress difference from commercial rheometers to quantify the values of both $\Psi_{1,0}$ and $\Psi_{2,0}$ for a series of polymer solutions. Furthermore, by retaining the oft-neglected inertial terms, we show that the ``climbing constant'' $\hat{\beta}=0.5\Psi_{1,0}+2\Psi_{2,0}$ can be measured even when the fluids, in fact, experience rod descending. A climbing condition derived by considering the competition between elasticity and inertial effects accurately predicts whether a fluid will undergo rod-climbing or rod-descending. Our results suggest a more general description, ``rotating rod rheometry'' instead of ``rod-climbing rheometry,'' to be more apt and less restrictive. The analysis and observations presented in this study establish rotating rod rheometry as a prime candidate for measuring normal stress differences in polymeric fluids at low shear rates that are often below commercial rheometers' sensitivity limits.
 

\end{abstract}

\maketitle

\section{Introduction}\label{sec:Intro}

The knowledge of a fluid's rheological properties is an essential prerequisite for predicting the flow of a complex fluid in any desired application. A simple steady shear flow measurement is enough for generalized Newtonian fluids as the shear-rate$-$dependent viscosity $\eta(\dot{\gamma})$ is the only rheological property required to resolve the flow dynamics in a specific geometry of interest. However, the presence of fluid elasticity requires using multiple deformation protocols to build a thorough understanding of the materials' rheological properties. Simple steady homogeneous shear flow, which is the most widely used test protocol, provides (in principle) quantitative information about the shear-rate dependence of three independent materials functions, viz., the viscosity $\eta(\dot{\gamma})$, the first normal stress difference $N_1(\dot{\gamma})$ and the second normal stress difference $N_2(\dot{\gamma})$. These normal stress differences are identically zero in Newtonian fluids. They are associated with nonlinear viscoelastic effects and hence, are negligibly small in linear viscoelastic measurements \cite{bird1987dynamics}. Their first appearance comes as a second-order effect in the shear rate, represented using the first ($\Psi_1$) and the second ($\Psi_2$) normal stress coefficients such that $N_1(\dot{\gamma})=\Psi_1\dot{\gamma}^2$ and $N_2(\dot{\gamma})=\Psi_2\dot{\gamma}^2$, respectively \cite{barnes1989introduction, larson2013constitutive}. However, due to limitations on the torque and axial force transducer sensitivities of commercial rheometers, it is typically only possible to measure the material functions over a limited range of shear rate values, which can be determined \textit{a priori} from the transducer sensitivity limits \cite{ewoldt2015experimental}.

In a strongly non-Newtonian fluid, $N_1$ can be comparable or even larger than the shear stress, $\sigma$, at high shear rates. Consequently, $N_1$ can typically be measured for many complex fluid systems using the very sensitive force re-balance transducer technology, which is now available in many commercial rheometers. Using a cone-and-plate (CP) geometry with a radius $R$ and cone angle $\theta_0$, the first normal stress difference $N_1(\dot{\gamma})$ can be measured directly from the axial force $F_{CP}$ acting on either the cone or the plate using \cite{barnes1989introduction}: 
\begin{equation}\label{eq:eq1}
    N_1(\dot{\gamma})=2F_{CP}(\dot{\gamma})/{\pi}R^2,
\end{equation}
where $\dot{\gamma}=\Omega/\theta_0$ and $\Omega$ is the rotational speed of the conical fixture. At low shear rates, in a simple fluid with fading memory\cite{bird1987dynamics}, $N_1$ is expected to vary quadratically through the analytical relationship
\begin{equation}\label{eq:eq2}
 \lim_{\dot{\gamma}\to 0}\frac{N_1}{\dot{\gamma}^2}=\Psi_1|_{\dot{\gamma} \to 0}=\Psi_{1,0}=\lim_{\omega \to 0}\frac{2G'(\omega)}{\omega^2}.
\end{equation}
Here the storage modulus $G'(\omega)$ can be measured with relatively high accuracy in small amplitude oscillatory shear (SAOS) flow at an oscillatory frequency $\omega$. The subscript $_0$ (e.g., $\Psi_{1,0}$) on a rheological property such as $\Psi_1$ denotes its value in the limit of a zero shear rate. 

On the other hand, the measurement of $N_2$ poses many difficulties compared to $\eta$ or $N_1$, resulting in far less attention to its accurate determination. Over the years, many techniques have been proposed to determine $N_2$ \cite{maklad2021review} experimentally. The most widespread approach is to use a cone-and-plate geometry for the direct measurement of $N_1(\dot{\gamma})$ using eq.~\ref{eq:eq1} in conjunction with another technique that measures a combination of $N_1$ and $N_2$. The complementary techniques are discussed in great detail in the review by \citet{maklad2021review} and include parallel-plate (PP) thrust measurement, offset cone-and-plate fixtures with distance adjustment, a cone-and-plate geometry with pressure distribution measurement, cone-and-partitioned plate, plate-and-ring geometry, and cone-and-ring geometry. Then an estimate of $N_2$ can be obtained by an appropriate subtraction of the two independent experimental measurements. However, because $N_2$ is often a small percentage of the value of $N_1$, such approaches are fraught with experimental difficulties.

Although the determination of $N_2$ using the combination of CP thrust and any of the supplementary measurements seems straightforward, many practical challenges arise, e.g., reduced values of the measured thrust due to inertia and/or secondary flows \cite{zilz2014serpentine}, amplified uncertainty due to subtracting two nearly equal values of the measured normal stress differences, differentiation of experimental data concerning various other parameters, or building a much more complicated experimental setup to measure pressure gradients (or several thrust measurements) directly\cite{maklad2021review}. However, the most commonly encountered limitation in measuring $N_1$ or $N_2$ is that the torque or thrust measuring transducers have minimum sensitivity limits below which they cannot detect the forces or torques exerted by a complex fluid under shear. For example, the state-of-the-art ARES-G2 rheometer (TA instruments) has a practical lower sensitivity limit of 0.001 N (0.1 gm-force) for the axial force $F_{CP}$ in Eq.~\ref{eq:eq1}. So, a 40 mm CP geometry, for instance, will not be sensitive to a first normal stress difference below approximately $N_1 \lessapprox 1.6$ Pa.s$^2$. Practically, this means that quantitative estimation of the asymptotic quadratic behavior of $N_1$ and $N_2$ (or equivalently the first and second normal stress coefficients $\Psi_1|_{\dot{\gamma} \to 0}= \Psi_{1,0}$ and $\Psi_2|_{\dot{\gamma} \to 0}=\Psi_{2,0}$) cannot be achieved using the above-mentioned techniques for many complex fluids of interest such as polymer solutions or concentrated suspensions. 

It is well known that the normal stress differences can also lead to secondary flows due to the breaking of axisymmetry, for instance, in pipes of non-circular cross-sections. These secondary flows arise because normal stress differences lead to tensile (or compressive) stresses acting along the streamlines and vortex lines in the flow (depending on the signs of $N_1$ and $N_2$). Similarly, non-zero contributions to the total stress acting on a deformable or free surface (for instance, flow down an open inclined trough or around a rotating rod, also lead to secondary flows). In such flows, the deformable shear-free surface acts as a very sensitive pressure ``gauge,” and hence, is a popular way to demonstrate visually and unequivocally the presence of normal stress differences in complex fluids. With careful work, the tilted trough (TT) experiment has been used to quantify $N_2$ by measuring the free surface deflection in the steady shear flow generated by gravity when the open trough is inclined at an angle \cite{wineman1966slow, tanner1970some}. The tilted trough technique offers great potential but requires a dedicated experimental facility, large volumes of fluid, and a complicated data-processing technique to extract $N_2$. In addition, only a narrow range of shear stresses can be probed, making it difficult to accurately determine $\Psi_{2,0}$ using the TT \cite{maklad2021review}. 

To overcome the above-mentioned limitations in reliably measuring $\Psi_{1,0}$ and $\Psi_{2,0}$ in the present study, we revisit the well-known rod-climbing effect. Following the original pioneering work by D.D. Joseph and co-workers on the rod-climbing rheometer \cite{,joseph1973free1,joseph1973free2,beavers1980free3,beavers1975rotating}, measurements of the climbing height and how it varies with rotation rate have been used to estimate $\Psi_{2,0}$ in polyisobutylene (PIB) Boger fluids with the same components but slightly different PIB concentrations (0.24 wt.\,\% \cite{magda1991second} and 0.1 wt.\,\% \cite{hu1990climbing}). However, this flow configuration has not received much attention in the ensuing three decades following these earlier studies, mainly because a reliable reconciliation of rod-climbing experiments with material functions measured in modern-day rheometers is still lacking. To this end, we present a protocol to robustly measure both $\Psi_{1,0}$ and $\Psi_{2,0}$ using a combination of rod-climbing, normal force measurements in steady shear, and SAOS measurements. The results presented here show that rod-climbing data can serve as an inexpensive supplement to data from commercial rheometers to enable measurements of both the first and second normal stress difference of a complex fluid in the low shear rate limit, which is typically beyond the sensitivity limits of commercial rheometers.

\begin{figure}
    \centering
    \includegraphics[width=0.45\textwidth]{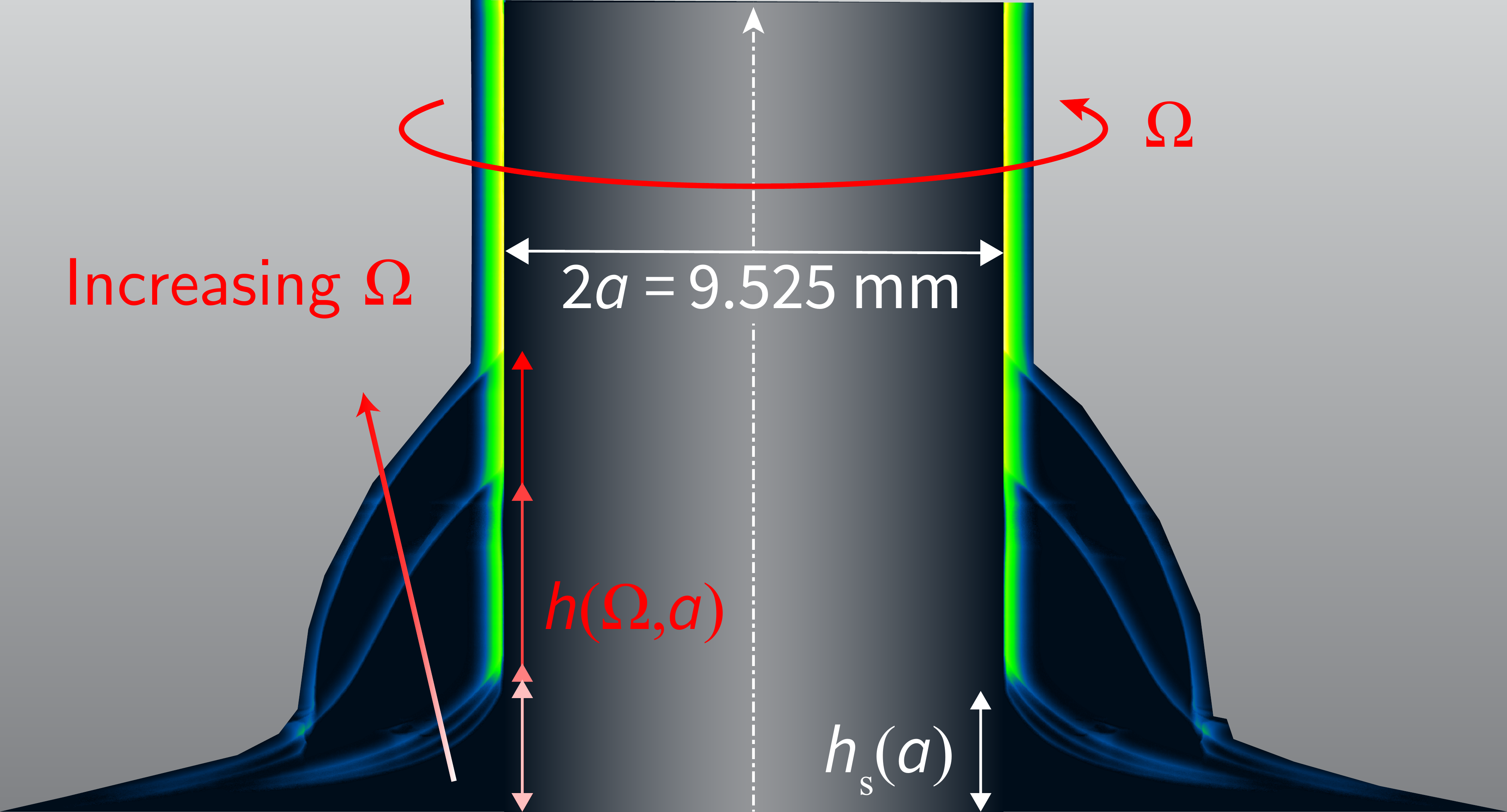}
    \caption{The rod-climbing or ``Weissenberg'' effect \cite{weissenberg1947non}. The free surface of an elastic fluid climbs a thin rotating rod with a radius $a$ and an angular velocity $\Omega$. The interface shape and the climbing height ($\Delta{h}(\Omega,a)$) compared to the static rise near the rod ($h_s(a)$) is primarily determined by the rate of rotation, the normal stresses in the fluid, along with surface tension and inertia.}
    \label{fig:fig1}
\end{figure}

\section{The climbing height in the ``rod-climbing'' experiment}
The ``rod climbing” or Weissenberg effect \cite{weissenberg1947non} is illustrated in Fig.~\ref{fig:fig1}, where the free surface of a fluid climbs a thin rotating rod and serves as an indisputable
experiment demonstrating the presence of non-linear elasticity in polymeric fluids. The presence of normal stresses in a fluid under shear leads to the idea of a streamwise or `hoop' stress in a fluid experiencing a torsional shear flow around a thin rotating rod. These hoop stresses pull fluid elements radially inward toward the rotating rod. As a result of this secondary flow, the deformable free surface near the rod ascends to a height at which the additional hydrostatic pressure pushing the fluid downwards and outwards exactly balances the hoop stress pulling the fluid inwards. The interface shape $h(\Omega,r,\alpha)$ and the climbing height $h(\Omega, a,\alpha)$, where $\Omega$ is the rod rotation speed, $a$ is the rod radius, and $\alpha$ is the contact angle, generally depend on the rod rotation rate and the fluid elasticity, as quantified by the two normal stress differences, surface tension, and inertia. The functional dependence can be determined by solving the governing equations of motion using a domain perturbation technique in the low rotation rate limit for a second-order fluid \cite{joseph1973free1,joseph1973free2,beavers1980free3}. A detailed derivation is presented in Appendix A. The final solution (Eq.~\ref{eq:eq21} in Appendix A) for the interface shape $h(\Omega,r,\alpha)$ gives the following mathematical relationship between the climbing height $h(\Omega, a,\alpha)$ and the material functions characterizing the fluid in terms of a specific combination of both normal stress differences called the ``climbing constant'' $\hat{\beta}=0.5\Psi_{1,0}+2\Psi_{2,0}$ \cite{joseph1973free2}:
\begin{equation}
\begin{aligned}\label{eq:eq3}
    h(\Omega,a,\alpha) = h_s(a,\alpha) + \frac{a}{2\left( \Gamma\rho{g} \right)^{1/2}}\left[ \frac{4\hat{\beta}}{4+\sqrt{Bo}}-\frac{\rho{a}^2}{2+\sqrt{Bo}} \right]\Omega^2 \\
    + O(\Omega^2\alpha+\Omega^4).
\end{aligned}
\end{equation}
Here $h_s(a,\alpha)$ is the static climbing height arising from capillarity effects (in a fluid with surface tension $\Gamma$ and contact angle $\alpha$). $\rho$ is the fluid density, $a$ is the rod radius which rotates with an angular speed $\Omega$, and $Bo={\rho}ga^2/\Gamma$ is the Bond number with $g$ being the acceleration due to gravity. Eq.~\ref{eq:eq3} establishes the foundation of rod-climbing rheometry \cite{maklad2021review}. 


\begin{figure*}[h]
    \centering
    \includegraphics[width=0.8\textwidth]{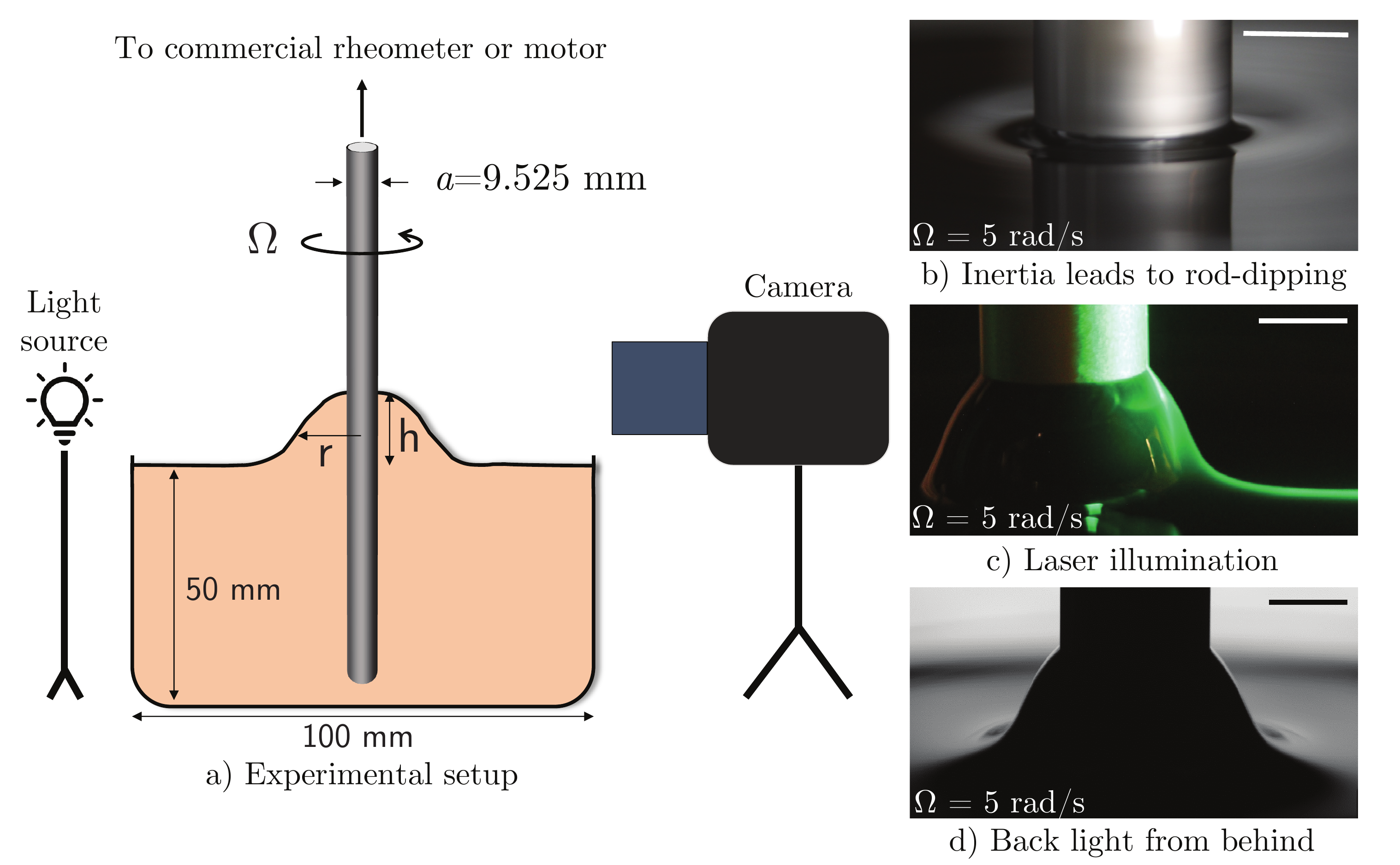}
    \caption{a) The rod-climbing rheometer includes a cylindrical beaker filled with the fluid of interest. A thin rod of radius $a$ is submerged with its axis aligned with the axis of the beaker. As the rod rotates with an angular velocity $\Omega$, the fluid interface climbs (or descends) if the climbing condition [see Sec.~\ref{sec:ClimbingCond} Fig.~\ref{fig:climbingcond}] is satisfied (not satisfied). The fluid is illuminated with a strong background light, and still images of the interface are captured using a digital camera (Nikon EOS 7D DSLR). b) In a weakly elastic fluid, dominant inertial effects compared to the fluid elasticity result in a local dip in the free surface that we define as ``rod-descending''. In a more strongly elastic fluid, rod-climbing due to large normal stress differences can be visualized; for example, by c) shining a laser sheet in a plane perpendicular to the camera and d) illuminating the fluid using a strong backlight. The bulk of the measurements in this work are performed using this latter illumination setup. Scale bars in (b)-(d) are all 5 mm.}
    \label{fig:setup}
\end{figure*}

\section{Methods and materials}\label{sec:method}

The schematic of our experimental setup is presented in Fig.~\ref{fig:setup}. A stress-controlled rheometer (TA instrument's AR-G2 Magnetic Bearing Rheometer) was modified to function as a rotating-rod rheometer. A precisely machined hollow steel tube of known diameter $2a=9.525$ mm was attached concentrically to an 8 mm parallel plate geometry to serve as a thin cylindrical rod that can be rotated in a fluid reservoir. The polymeric fluid (see description below) was contained in a cylindrical glass beaker of diameter 100 mm and depth 50 mm. The original study by \citet{joseph1973free2} recommends the beaker-to-rod diameter ratio to be at least ten so that edge effects on the climbing height measured in the low rotation speed regime are minimal\cite{hu1990climbing, magda1991second}. The rods were fully immersed in the beaker to a depth of $10a \approx 45$ mm. It has been shown that the immersion depth of the rod into the fluid does not affect the rod-climbing height \cite{magda1991second}. The position of the beaker was manually adjusted to align it concentrically with the rotation axis of the rotating rod. Using the AR-G2 rheometer motor to impose the rotation rate allowed us to control the rod rotating speed $\Omega$ accurately. Photographs taken with a Nikon EOS 7D DSLR camera were used to capture the free surface shape and the climbing height at various rotational speeds. The photographs were used to measure the climbing heights of the fluid-free surface near the rod $h(\Omega, a,\alpha)$. The photographs capture the interface with a spatial resolution of 0.05 mm/pixel, which can be determined from the magnification of the lens and the pixel resolution of the captured images. 

Eq.~\ref{eq:eq3} for the rod-climbing height at the rod surface ($r=a$) was derived assuming a semi-infinite fluid container and a domain perturbation approach (See Appendix A for a detailed derivation). All the higher-order terms in Eq.~\ref{eq:eq3} involve computing secondary motions from complex boundary value problems. However, it has been shown that computing these higher-order terms is not necessary at low rotation rates and the expression
\begin{equation}\label{eq:eq5}
\begin{aligned}
    \Delta{h}(\Omega,a) &= h(\Omega,a,\alpha) - h_s(a,\alpha) \\
    &\approx \frac{a}{2\left( \Gamma\rho{g} \right)^{1/2}}\left[ \frac{4\hat{\beta}}{4+\sqrt{Bo}}-\frac{\rho{a}^2}{2+\sqrt{Bo}} \right]\Omega^2 
    \end{aligned}
\end{equation}
is a good approximation for the changes in the free surface height in the small rotation speed limit \cite{joseph1973free1, joseph1973free2}. This means the change (or perturbation) in the climbing height $\Delta{h}(\Omega,a)=h(\Omega,a,\alpha)-h_s(a,\alpha)$ is approximately independent of the contact angle $\alpha$ in the small $\Omega$ limit and scales linearly with $\Omega^2$. 
From the rod-climbing experiment (Fig.~\ref{fig:setup}), we measure the change in the climbing height due to rod rotation $\Delta{h}(\Omega, a)$ compared to the static rise height $h_s(a,\alpha)$ and plot it as a function of the imposed rotation speed. From this plot of $\Delta{h}(\Omega,a)$ vs. $\Omega^2$, we compute the slope $\textrm{d}\Delta{h}/\textrm{d}\Omega^2$ in the quadratic regime corresponding to low values of $\Omega^2$ and equate it to the theoretical slope
\begin{equation}\label{eq:eq6}
    \frac{\textrm{d}\Delta{h}(\Omega,a)}{\textrm{d}\Omega^2} \approx \frac{a}{2\left( \Gamma\rho{g} \right)^{1/2}}\left[ \frac{4\hat{\beta}}{4+\sqrt{Bo}}-\frac{\rho{a}^2}{2+\sqrt{Bo}} \right] 
\end{equation}
obtained from Eq.~\ref{eq:eq5}. If the fluid density and surface tension are known, one can compute the climbing constant $\hat{\beta}=0.5\Psi_{1,0}+2\Psi_{2,0}$ from the slope of the data. This calculated value of $\hat{\beta}$ can then be used in conjunction with an independent measurement of $\Psi_{1,0}$ from a complementary method to calculate $\Psi_{2,0}$ as
\begin{equation}\label{eq:eq7}
    \Psi_{2,0}=\frac{1}{2}\hat{\beta}-\frac{1}{4}\Psi_{1,0}=-\frac{\Psi_{1,0}}{4}\left( 1- \frac{2\hat{\beta}}{\Psi_{1,0}} \right).
\end{equation}
As discussed in the Introduction, direct measurements of the axial force in a CP geometry used to measure $N_1$ using Eq.~\ref{eq:eq1} cannot probe very small shear rates because the normal force signal exerted by the fluid is often weaker than the lower sensitivity limits of the force transducer. Hence, measuring $G'(\omega)$ (with relatively high accuracy) through SAOS deformations in a concentric cylinders geometry and using the asymptotic limit in Eq.~\ref{eq:eq2}, i.e., $\Psi_{1,0} \simeq \lim_{\omega \to 0}\frac{2G'}{\omega^2}$ is a practical and superior solution, as will be shown later in Sec.~\ref{sec:Results}.  

Further insights into the competition between the elastic effects (i.e., the first term in the square brackets in Eq.\ref{eq:eq6} $-$ denoted Term I), which encourage rod-climbing, and the inertial effects, which encourage a decrease in the height (the second term in the square brackets in Eq.\ref{eq:eq6} $-$ denoted Term II), can be obtained from Eq.~\ref{eq:eq6}. These insights aid in extending the usefulness of the rod-climbing rheometer beyond the moderately viscous and elastic fluids tested decades ago \cite{hu1990climbing,magda1991second, magda1994concentrated}. 
To achieve this expansion in utility, we do not make any simplification to Eq.~\ref{eq:eq6} and, unlike previous studies \cite{hu1990climbing,magda1991second, magda1994concentrated}, retain the contribution due to inertial effects (Term II). From Eq.~\ref{eq:eq6}, one can conclude that weakly elastic fluids with relatively smaller values of $\hat{\beta}$ can undergo rod-descending (i.e., Term I $<$ Term II in Eq.\ref{eq:eq6}), that is, the climbing height decreases with increasing $\Omega$ from the initial static rise $h_s(a,\alpha)$ due to the dominance of inertial effects. However, the linear relationship in $\Delta{h}(\Omega,a)$ vs. $\Omega^2$ in the low rotation rate limit is still valid and can be used to measure $\hat{\beta}$ (and consequently $\Psi_{2,0}$) even in weakly viscoelastic fluids, e.g., dilute or semi-dilute polymer solutions. We test this hypothesis in Sec.~\ref{sec:Results}. In the extreme case of a Newtonian fluid when $\hat{\beta}=0$ we expect
\begin{equation}\label{eq:eq8}
    \frac{\textrm{d}\Delta{h}(\Omega,a)}{\textrm{d}\Omega^2} \approx \frac{a}{2\left( \Gamma\rho{g} \right)^{1/2}}\left[-\frac{\rho{a}^2}{2+\sqrt{Bo}} \right], 
\end{equation}
which sets the Newtonian inertial rod-dipping limit on the slope of the $\Delta{h}(\Omega,a)$ vs. $\Omega^2$ plot (in the low $\Omega$ limit). Any deviation of $\textrm{d}\Delta{h}/\textrm{d}\Omega^2$ from Eq.~\ref{eq:eq8}, denotes a presence of a non-zero climbing constant $\hat{\beta}$, and consequently, the presence of finite normal stress differences. Eq.~\ref{eq:eq6} can also be utilized to derive a ``climbing condition'' in terms of two dimensionless quantities: a dimensionless normal stress difference ratio $\psi_0=-\Psi_{2,0}/\Psi_{1,0}$ and the inertioelastic quantity $\rho{a^2}/\Psi_{1,0}$; both of which are independent of the flow kinematics in the problem. We defer a more detailed discussion on this climbing condition to Sec.~\ref{sec:ClimbingCond}.

\begin{table}
\small
  \caption{\ PIB polymer solution properties.}
  \label{tab:tab1}
\begin{tabular*}{0.48\textwidth}{@{\extracolsep{\fill}}c c c c c}
    \hline
    C (wt.\,\%) & $\rho$ (kg/m$^3$) & $\Gamma$ (mN/m) & $\eta_s$ (mPa.s) & C$^*$ (wt.\,\%)\\
    \hline
    0.30 $-$ 3.00 & 873.1 & 29.7 & 18.07 & 0.23 \\
    \hline
  \end{tabular*}
\end{table}

For this study, we use polymeric solutions of polyisobutylene (PIB) (Molecular weight $\approx 10^6$ g/mol) dissolved in a paraffinic oil (Lubrizol Inc.). We perform all our measurements at a constant temperature $T=20$ $^\circ$C. The Newtonian solvent oil has a steady state viscosity $\eta_s=18.07$ mPa.s at $20$ $^\circ$C. The polymer intrinsic viscosity is measured to be $\left[ \eta \right] = 3.69$ dL/g, and this can be used to estimate the critical overlap concentration of the polymer solute C$^* \simeq 0.77/\left[ \eta \right]$\cite{graessley1980polymer} $= 0.23$ wt.\,\%. The solutions were all measured to have a constant density $\rho=873.1$ kg/m$^3$ and surface tension $\Gamma=29.7$ mN/m. The material properties of the test fluids used are summarized in Table~\ref{tab:tab1}. We vary the dissolved concentration of polymer in the solution to change the viscoelastic properties. We work with three semi-dilute solutions (C $>$ C$^*$): 3 wt.\,\%, 2 wt.\,\%, and 1 wt.\,\%, respectively, and one close to C$^*$: 0.3 wt.\,\%. These polymeric solutions are all shear-thinning, with their viscoelasticity decreasing at lower concentrations, as shown in the next section.  

\section{Results and discussion}\label{sec:Results}

The rotating rod rheometry protocol involves 
\begin{enumerate}
    \item determining the value of $\Psi_{1,0}$ from SAOS data obtained over a range of temperatures to construct a time-Temperature superposition (tTS) master curve and then using the asymptotic result obtained from simple fluid theory $\Psi_{1,0} = \lim_{\omega\to0}2G'/\omega^2$ (Sec.~\ref{sec:SAOS}),
    \item calculating the climbing constant $\hat{\beta}=0.5\Psi_{1,0}+2\Psi_{2,0}$ by measuring the surface deflection and determining the slope $\textrm{d}\Delta{h}/\textrm{d}\Omega^2$ of perturbations to the static interface $\Delta{h}(\Omega,a)$ vs. $\Omega^2$ for small $\Omega$. This value is then equated to the theoretical result (Eq.~\ref{eq:eq6}) (Sec.~\ref{sec:RodClimbing}), and
    \item calculating the second normal stress coefficient using the relationship $\Psi_{2,0} = (\hat{\beta}-0.5\Psi_{1,0})/2$ (Sec.~\ref{sec:Reconcile}).
\end{enumerate}
We first present the results of applying this protocol to the four PIB-based fluids described above. In addition, we derive a modified ``climbing condition'' (Sec.~\ref{sec:ClimbingCond}) and present observations which support the use of the rotating rod experiment to probe $\Psi_{2,0}$ even in fluids that exhibit a local dip in the free surface that we define as ``rod-descending'' (Sec.~\ref{sec:Boger}).  

\subsection{$\Psi_{1,0}$ measurements using small amplitude oscillatory shear}\label{sec:SAOS}

\begin{figure}[t!]
  \centering
    \includegraphics[width=0.47\textwidth]{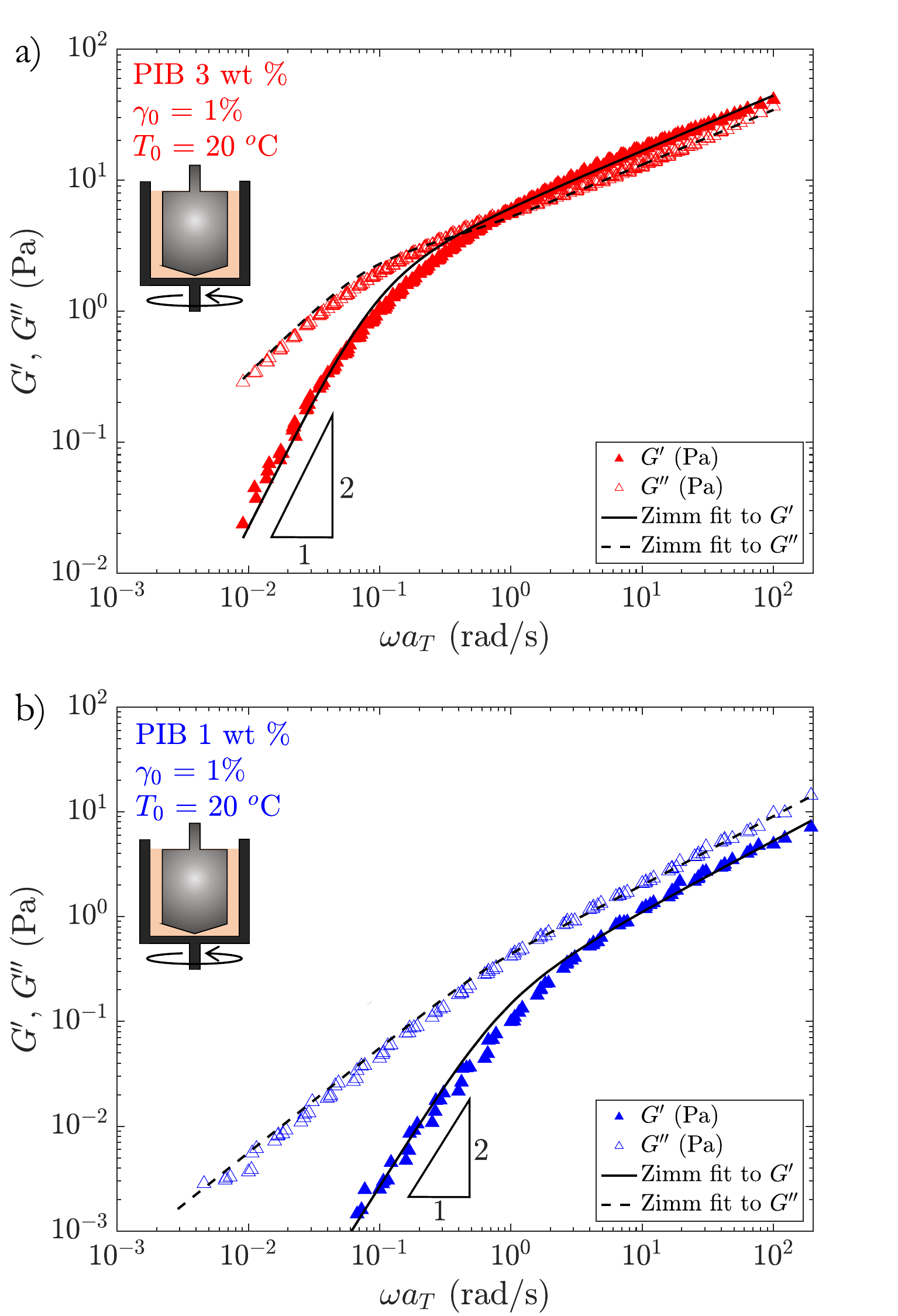}
   \caption{\label{fig:SAOSzimm}Small amplitude oscillatory shear (SAOS) measurements at a strain amplitude $\gamma_0=1$ \% of a) 3 wt.\,\% and b) 1 wt.\,\% PIB solutions used in this study. Time-temperature superposition using a reference temperature $T_0=20$ $^{\circ}C$ is employed to construct a master curve with a lateral shift factor $a_T$ giving the reduced frequency $\omega_{r} = a_T{\omega}$. The generalized Rouse-Zimm model (Eq.~\ref{eq:eq9}) does an excellent job of fitting the SAOS data, and the corresponding fits are shown using black lines. The fitting parameter values for the generalized Rouse-Zimm model are tabulated in Table~\ref{tab:zimmfit}.}
\end{figure}

Fig.~\ref{fig:SAOSzimm} shows the results obtained from a small amplitude oscillatory shear (SAOS) flow experiments using a concentric cylinder geometry for the 3 wt.\,\% and 1 wt.\,\% solutions. We perform time-Temperature superposition (tTS) to construct a master curve of the storage and loss moduli as functions of the reduced oscillation frequency $\omega_r=a_T\omega$, denoted $G'(\omega_r)$ and $G''(\omega_r)$, respectively. This allows us to extend the range of measurements to sufficiently low frequencies to observe the terminal scaling expected. Here $\omega$ is the oscillatory frequency, and $a_T$ is the temperature-dependent horizontal shift factor. For the limited range of temperatures 10 $^\circ$C $\leq T \leq 80$ $^\circ$C studied here, we find the vertical shift factor $b_T \approx 1$ for these PIB solutions, and there is no need to shift the $G'(\omega_r)$ and $G''(\omega_r)$ data vertically. In addition, we observe that even though the solutions are in the semi-dilute regime, the generalized Rouse-Zimm model expressed in the form \cite{rubinstein2003polymer}
\begin{subequations}\label{eq:eq9}
\begin{equation}\label{eq:eq9a}
    G'(\omega_{r}) = G_c\frac{\omega_{r}\tau_Z\textrm{sin}\left[ \chi \textrm{atan} (\omega_{r} \tau_Z) \right]}{\left[ (1+(\omega_{r} \tau_Z)^2) \right]^{\chi/2}},
\end{equation}
\begin{equation}\label{eq:eq9b}
    G''(\omega_{r}) = G_c\frac{\omega_{r}\tau_Z\textrm{cos}\left[ \chi \textrm{atan} (\omega_{r} \tau_Z) \right]}{\left[ (1+(\omega_{r} \tau_Z)^2) \right]^{\chi/2}}
\end{equation}
\end{subequations}
does a good job of fitting the SAOS data, as shown in Fig.~\ref{fig:SAOSzimm}. Here the fitting parameters are the characteristic modulus $G_c$, the Zimm relaxation time $\tau_Z$, and $\chi=(1-1/3\nu)$ with $\nu$ being the solvent quality exponent. The best-fit parameter values for the generalized Rouse-Zimm model are tabulated in Table~\ref{tab:zimmfit}. 

\begin{table}[h!]
\small
  \caption{\ Generalized Rouse-Zimm model fit parameters for the linear viscoelastic properties obtained from master curves of the small amplitude oscillatory shear (SAOS) flow data for various PIB solutions, as well as the first normal stress coefficient obtained from Eq.~\ref{eq:eq10}.}
  \label{tab:zimmfit}
  \begin{tabular*}{0.48\textwidth}{@{\extracolsep{\fill}}c c c c c c c}
   \hline
    C (wt.\,\%) & $G_c$ (Pa) & $\tau_z$ (s) & $\chi$ & $\nu$ & $:$ & $\Psi_{1,0}$ (Pa.s$^2$)\\
    \hline
    3.00 & 2.88 & 11.62 & 0.579 & 0.791 & $:$ & 534.0 \\
    2.00 & 1.03 & 7.01 & 0.514  & 0.686 & $:$ & 051.9 \\ 
    1.00 & 0.39 & 1.43 & 0.337 & 0.503 & $:$ & 0.527 \\
    0.30 & 0.20 & 0.52 & 0.189 & 0.411 & $:$ & 0.021 \\
   \hline
  \end{tabular*}
\end{table}

We can now use Eq.~\ref{eq:eq2} to evaluate the expected value of the first normal stress difference coefficient $\Psi_{1,0}$ in the zero shear limit. Using the generalized Rouse-Zimm model (Eq.~\ref{eq:eq9a}) and the expectation from simple fluid theory (Eq.~\ref{eq:eq2}) we obtain, 
\begin{equation}\label{eq:eq10}
    \Psi_{1,0} = 2G_c\tau_Z^2\chi,
\end{equation}
which allows us to calculate $\Psi_{1,0}$ from the linear viscoelastic master curve. The calculated values for all the solutions used in this study are provided in Table~\ref{tab:zimmfit}. However, it should be noted that the asymptotic value of $\Psi_{1,0}$ as defined in Eq.~\ref{eq:eq2}, can also be calculated solely from a master curve of the SAOS data by a careful regression analysis of the empirical data. This is especially useful if a statistically good fit with available models, such as the Zimm model, is not possible; provided there is a discernible quadratic regime in which $G'(\omega_r) \sim \omega_r^2$ at low frequencies. Observing a clear quadratic scaling from the SAOS data at a fixed temperature can often be difficult for certain weakly elastic fluids, as sufficiently low frequencies might not be accessible due to rheometer sensitivity limits. Hence, performing time-temperature superposition can be useful, as well as using a constitutive model so that SAOS measurements can be robustly extrapolated into the region where $\tau_Z\omega_r \ll 1$ to calculate $\Psi_{1,0}$ accurately. Results in Table~\ref{tab:zimmfit} show that we can expect $\Psi_{1,0}$ to vary over four orders of magnitude by diluting the solutions over a factor of 10. Thus, varying the PIB concentration is expected to have an equivalent dramatic impact on the rod-climbing; indeed, we observe the same sensitivity level as discussed in the following subsection.

\begin{figure*}
    \centering
    \includegraphics[width=0.8\textwidth]{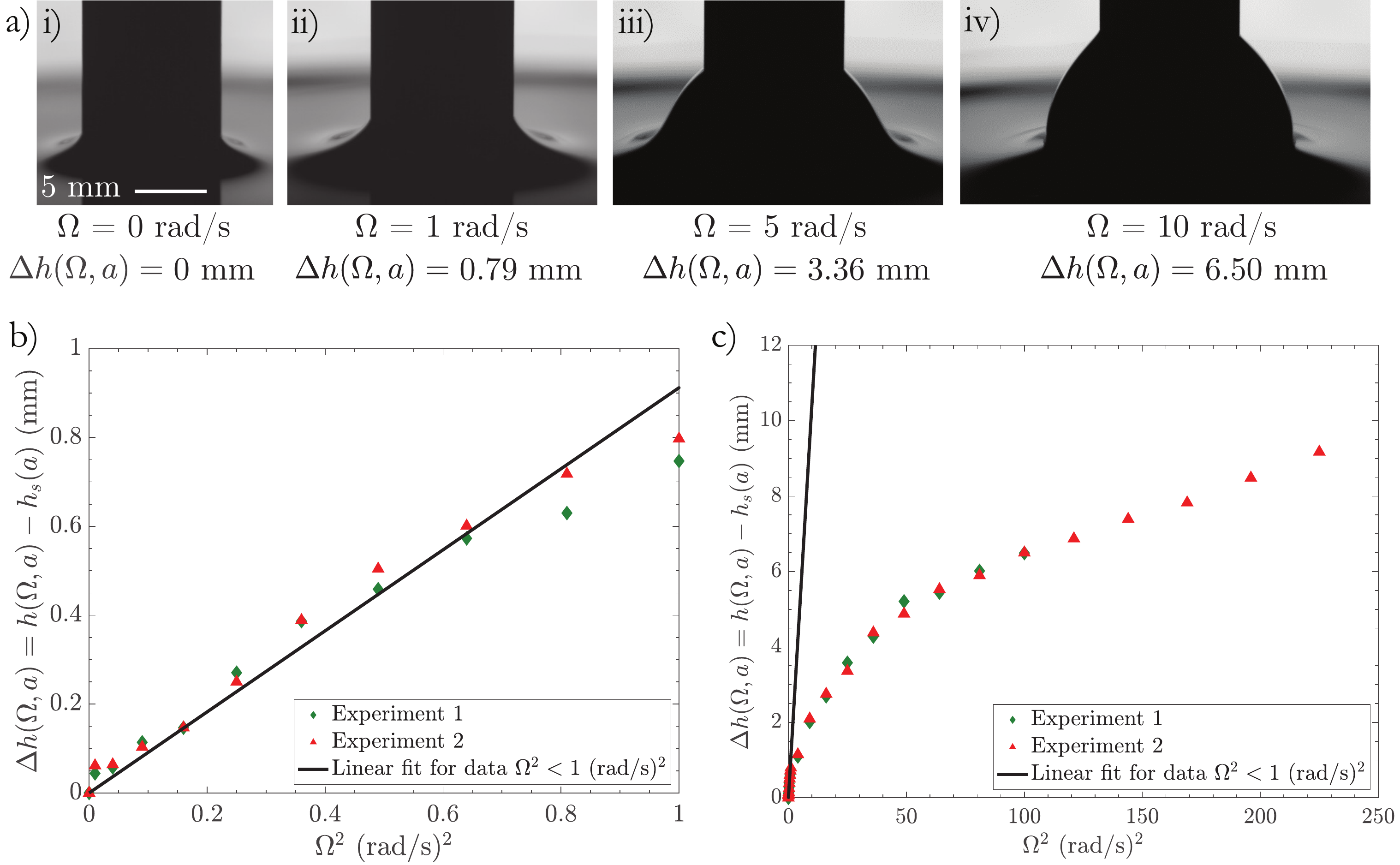}
    \caption{a) Still images illustrating rod-climbing in the 3 wt.\,\% PIB solution with increasing rod rotation speeds i) $\Omega=0$ rad/s, ii) $\Omega=1$ rad/s, iii) $\Omega=5$ rad/s, and $\Omega=10$ rad/s. The change in the interface height at the rotating rod ($\Delta{h}(\Omega,a)$) compared to the static rise $h_s(a)$ for 3 wt.\,\% PIB solution: b) In the low $\Omega$ regime ($\Omega < 1$ rad/s), $\Delta{h}$ varies linearly with $\Omega^2$ as predicted by theory (Eq.~\ref{eq:eq5}). The slope of the curve in this regime is used to calculate the climbing constant $\hat{\beta}$. c) At higher $\Omega$, the higher order terms in the perturbation expansion cannot be neglected and lead to secondary flows in the fluid bulge that result in the deviation in linearity from a plot of $\Delta{h}$ vs. $\Omega^2$.}
    \label{fig:Climbing3wt}
\end{figure*}

\subsection{Measurements of the climbing constant $\hat{\beta}$ using rod-climbing observations}\label{sec:RodClimbing}
We turn our attention to the rod-climbing experiments in this subsection. The experimental data and representative photographs of the 3 wt.\,\% solution undergoing rod-climbing are presented in Fig.~\ref{fig:Climbing3wt}. When $\Omega=0$, we observe a finite climbing height due to meniscus wetting as shown in Fig.~\ref{fig:Climbing3wt}a-i. This is the static climbing height $h_s(a,\alpha)$ as explained in Eq.~\ref{eq:eq5}. This meniscus height and shape can be adequately described by solving the Young-Laplace equation for the interface with the knowledge of the contact angle determined from the photograph, but this will not be pursued here as our focus is on measuring the \textit{change} in the climbing height $\Delta{h}(\Omega,a)$ due to rotation of the rod. As we rotate the rod with a rotation speed $\Omega$, $\Delta{h}(\Omega,a)$ is expected to initially be proportional to $\Omega^2$ at low rotation rates, as shown in Fig.~\ref{fig:Climbing3wt}b. However, it is difficult to predict \textit{a priori} the maximum allowable rotational speed of the rod (denoted $\Omega_{max}$) above which the experimental observations deviate from the quasi-linear relationship given by Eq.~\ref{eq:eq6}. In practice, one can estimate $\Omega_{max}$ from the experimental data \textit{a posteriori} as shown in Fig.~\ref{fig:Climbing3wt}c with the condition that the modified Froude number $Fr = \Omega^2L/g < 1$. Here $L$ is a characteristic length, which can be taken to be $L \approx \sqrt{g\textrm{d}\Delta{h}/\textrm{d}\Omega^2}$\cite{joseph1973free2}$\approx \frac{1}{2}\sqrt{a\hat{\beta}\sqrt{\frac{g}{\rho\Gamma}}}$ using Eq.~\ref{eq:eq6}. From the rod-climbing observations, using a conservative condition that $Fr < 1$,  we find $\Omega_{max} \approx$ 2, 6, 15, and 30 rad/s for the 3 wt.\,\%, 2 wt.\,\%, 1 wt.\,\%, and 0.3 wt.\,\% PIB solutions, respectively.

\begin{figure}
    \centering
    \includegraphics[width=0.47\textwidth]{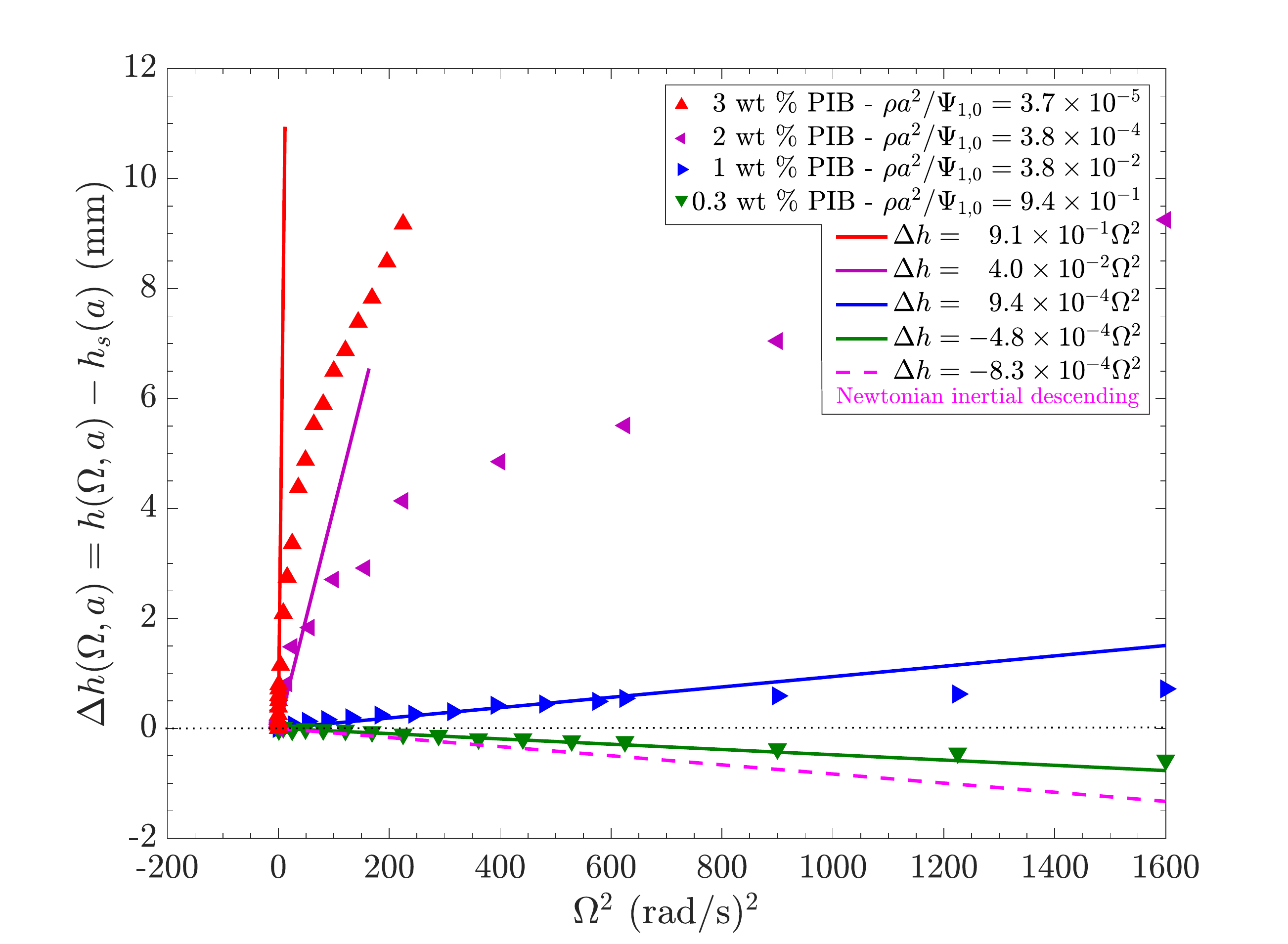}
    \caption{Change in the interface height at the rotating rod ($\Delta{h}(\Omega,a)$) compared to the static rise $h_s(a)$ for various PIB solutions utilized in this study. The 0.3 wt.\,\% solution does not satisfy the condition for climbing (Sec.~\ref{sec:ClimbingCond} Fig.~\ref{fig:climbingcond}); hence, the interface height decreases with the rotation rate. However, this rod-descending regime still varies linearly with $\Omega^2$ as predicted by the domain perturbation solution for a second-order fluid in the low rotational speed limit. The dashed line depicts the lower bound on rod-descending expected for a purely Newtonian fluid without any elasticity, which arises solely due to inertial effects. The $\Delta{h}$ vs. $\Omega^2$ curves of any fluid with finite non-zero normal stresses will lie above this lower bound. }
    \label{fig:allheight}
\end{figure}

The observed values of $\Delta{h}(\Omega, a)$ lie on a straight line when plotted against $\Omega^2$ at values $\Omega<\Omega_{max}$ as shown in Fig.~\ref{fig:Climbing3wt}b, c. The photographs of the interface shape presented in Fig.~\ref{fig:Climbing3wt}a-i, ii reveal that the interface shape $h(\Omega,r,\alpha)$ is concave for $\Omega \lesssim \Omega_{max}$. This shows that in the low $\Omega$ regime, increasing the rotation rate results in small perturbations to the static interface shape, which increases the interface height linearly with $\Omega^2$ as predicted by Eq.~\ref{eq:eq6}. These perturbations to the interface shape $\Delta{h}(\Omega,a)$ are positive if elastic effects dominate over inertial effects, i.e., if Term I $>$ Term II in Eq.~\ref{eq:eq6}. In other words, we should observe rod-\textit{climbing}, which is true for concentrated solutions with C = 3 wt.\,\%, 2 wt.\,\%, and 1 wt.\,\% as shown in Fig.~\ref{fig:allheight}. On the other hand, the perturbations to the static interface shape $\Delta{h}(\Omega,a)$ are negative if elastic effects are weaker than inertial effects, i.e., Term I $<$ Term II in Eq.~\ref{eq:eq6}. In other words, we should observe rod-\textit{descending}, which is true for the semi-dilute solution with C = 0.3 wt.\,\% as shown in Fig.~\ref{fig:allheight}. Thus, from Eq.~\ref{eq:eq6}, it is evident that we can predict whether a given complex fluid will undergo rod-climbing or descending if its material functions are known. We will return to this discussion in Sec.~\ref{sec:ClimbingCond}, where we analyze Eq.~\ref{eq:eq6} in further detail and derive a ``climbing condition'' to predict whether the interface of a given viscoelastic fluid will climb or descend a thin rotating rod immersed in it.   

For larger values of rotation speeds $\Omega^2>\Omega_{max}^2$ when rod-climbing is observed (Term I $>$ Term II in Eq.~\ref{eq:eq6}), the height increment $\Delta{h}(\Omega,a)$ increases non-linearly with $\Omega^2$ and the asymptotic relationship presented in Eq.~\ref{eq:eq6} is no longer valid. The interface shape changes from concave to convex in the large $\Omega$ regime as depicted in Fig.~\ref{fig:Climbing3wt}a-iii. Consequently, this transition from a concave to a convex interface can also be used as an \textit{in situ} condition to determine $\Omega_{max}$ more accurately (in conjunction with the criterion involving the modified Froude number), and this experimentally motivated approach is utilized in this study. With further increase in the rod rotation rate beyond $\Omega_{max}$, the interface shape assumes a rotating blob-like shape, which emerges distinctively from the larger pool of stationary or slowly rotating fluid at a point with a slope discontinuity as shown in Fig.~\ref{fig:Climbing3wt}a-iv. Eventually, at very high rotation speeds, this bolus of fluid becomes unstable with unsteady secondary motions resulting in a band of fluid rising up and down the rod in a wave-like manner. Further increases in $\Omega$ completely disrupt the climbing fluid blob into smaller pendant drops that are thrown radially outwards from the rod.    

From our rod-climbing experiments as well as previous studies \cite{joseph1973free2,beavers1975rotating,magda1991second}, we can conclude that there is an accessible range of small rotation rates $\Omega<\Omega_{max} \approx \sqrt{g\textrm{d}\Delta{h}/\textrm{d}\Omega^2}$ such that the second-order fluid approximation is valid and the climbing fluid interface height scales linearly with $\Omega^2$. Hence, we can equate the slope ($\textrm{d}\Delta{h}/\textrm{d}\Omega^2$)$_{exp}$ determined experimentally in the $\Omega<\Omega_{max}$ regime to the predictions of the second-order fluid theory in Eq.~\ref{eq:eq6} to evaluate the climbing constant 
\begin{equation}
\hat{\beta}_{exp} = \frac{4+\sqrt{Bo}}{4}\left[ \left(\frac{\textrm{d}\Delta{h}}{\textrm{d}\Omega^2}\right)_{exp}\frac{2\left( \Gamma\rho{g} \right)^{1/2}}{a} + \frac{\rho{a}^2}{2+\sqrt{Bo}} \right].          
\end{equation} 
The values of $\hat{\beta}_{exp}$ thus obtained from measurements for the various PIB solutions are presented in Table~\ref{tab:results}. We observe a significant reduction in the climbing constant $\hat{\beta}$ that varies over four orders of magnitude by diluting the PIB concentration from 3 wt.\,\% to 0.3 wt.\,\% as was anticipated in Sec.~\ref{sec:SAOS} from a similar dramatic four orders of magnitude of reduction in the $\Psi_{1,0}$ values.  

\begin{table}[h]
\small
  \caption{\ Measurements of the experimental climbing constant $\hat{\beta}_{exp}$ from the rod-climbing rheometry and $\Psi_{2,0}$ by combining rod-climbing rheometry with the SAOS measurements from conventional rheometer (TA instruments ARES-G2) for the various PIB polymer solutions utilized in this study.}
  \label{tab:results}
  \begin{tabular*}{0.48\textwidth}{@{\extracolsep{\fill}}c c c c c}
    \hline
    C (wt.\,\%) & $\hat{\beta}_{exp}$ (Pa.s$^2$) & $\psi_0=-\frac{\Psi_{2,0}}{\Psi_{1,0}}$ & $\Psi_{1,0}$ (Pa.s$^2$) & $\Psi_{2,0}$ (Pa.s$^2$) \\
    \hline
    3.00 & 10.03 & 0.241 & 534.0 & $-129.0$ \\
    2.00 & 0.710 & 0.243 & 051.9 & $-012.6$ \\ 
    1.00 & 0.018 & 0.233 & 0.527 & $-0.123$ \\
    0.30 & 0.002 & 0.205 & 0.021 & $-0.004$ \\
    \hline
  \end{tabular*}
\end{table}

\subsection{Reconciling rod-climbing measurements with SAOS to determine $\Psi_{2,0}$}\label{sec:Reconcile}
Reptation theory predicts that the normal stress difference ratio $\psi_0=-\Psi_{2,0}/\Psi_{1,0}$ for semi-dilute and concentrated entangled solutions has the value $\psi_0$ = 2/7 or 1/7 depending on whether the independent alignment assumption is made or not \cite{kimura1981polym}. The precise value of $\psi_0$ critically affects the level of rod climbing expected. Specifically, if $\psi_0=1/4$, then from Eq.~\ref{eq:eq7}, it is clear that the climbing constant is $\hat{\beta}=0$.

However, in dilute solutions without entanglement effects, one can expect $\psi_0=0$ as has been confirmed for some Boger fluids experimentally \cite{keentok1980measurement,magda1991second} and thus conclude that rod-climbing is driven primarily from the first normal stress difference. This becomes clear if we rearrange Eq.~\ref{eq:eq7} as
\begin{equation}\label{eq:eq7re}
    \Psi_{1,0}=\frac{2\hat{\beta}}{1+4\Psi_{2,0}/\Psi_{1,0}} = \frac{2\hat{\beta}}{1-4\psi_0}
\end{equation}
Then by substituting $\psi_0=0$ (or equivalently $\Psi_{2,0}=0$ in the first inequality of Eq.~\ref{eq:eq7re}, one obtains $\Psi_{1,0}=2\hat{\beta}$. In this limit from the rod-climbing measurements of $\hat{\beta}_{exp}$, one could calculate the expected value of the first normal stress difference coefficient to be $\Psi_{1,0} = 2\hat{\beta}_{exp}$. This \textit{a priori} estimate of $\Psi_{1,0}=2\hat{\beta}_{exp}$ can be interpreted as the \textit{lower bound} on $\Psi_{1,0}$ obtained from rod-climbing measurements alone because any finite positive $\psi_0$ value would result in a larger computed value of $\Psi_{1,0}$. In other words, in the absence of second normal stress effects in the fluid, the lower bound value of $\Psi_{1,0}=2\hat{\beta}_{exp}$ obtained above should be enough to achieve a rod-climbing height measured in the experiments. However, the presence of a non-zero second normal stress difference diminishes the rod-climbing abilities of the fluid, as a result of which, a higher value of $\Psi_{1,0}$ is required to achieve the climbing height given by the measured climbing constant $\hat{\beta}_{exp}$.

\begin{figure}[h!]
 \centering
 \includegraphics[width=0.45\textwidth]{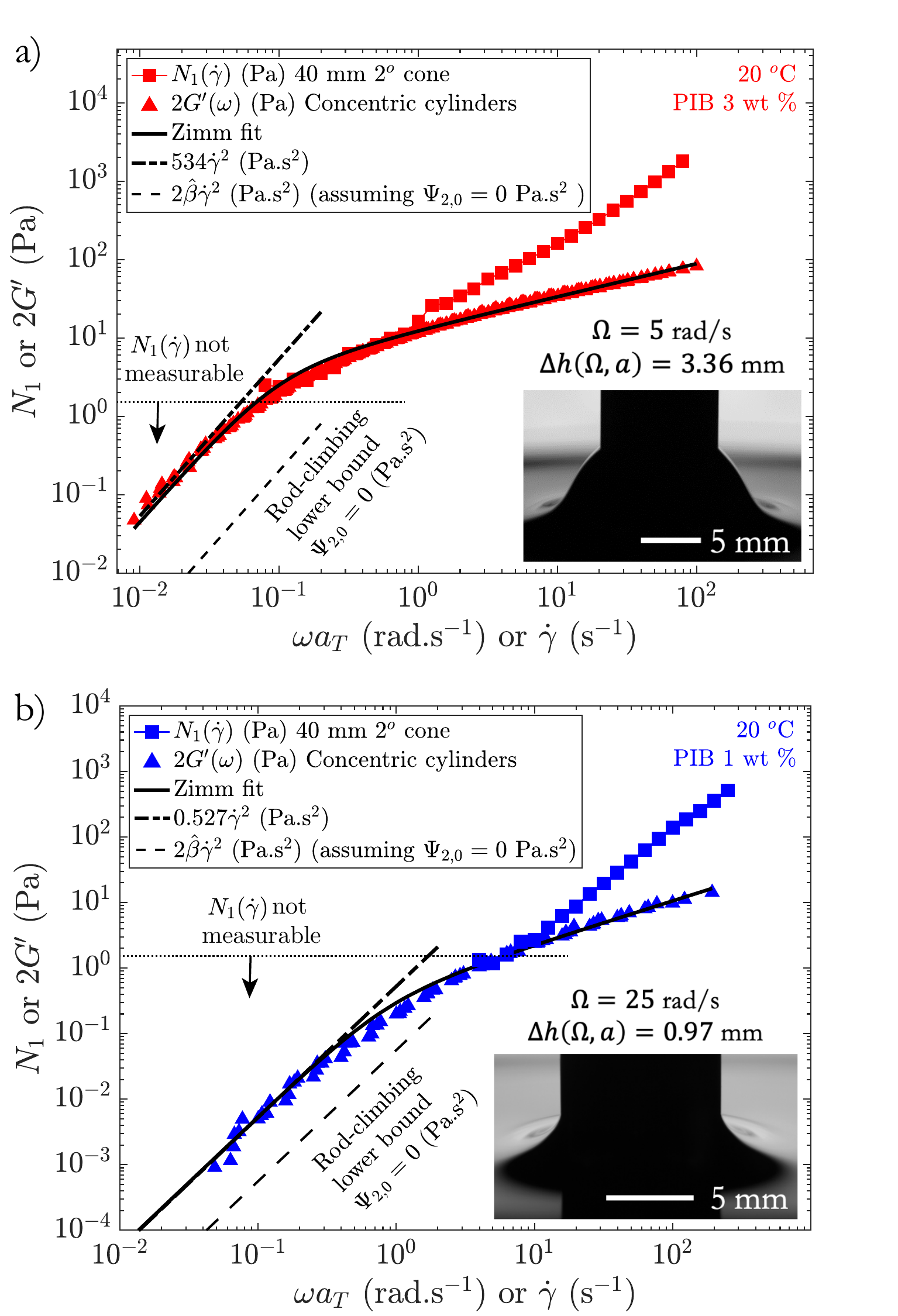}%
   \caption{\label{fig:reconciling}Reconciling rod-climbing measurements of normal stress differences measurements with conventional rheometry for: a) 3 wt.\,\% PIB solution, and b) 1 wt.\,\% PIB solution. $\Psi_{1,0}$ can be estimated from the SAOS data using the simple fluid asymptotic theory, which gives $\Psi_{1,0}=\lim_{\omega \to 0}2G'/\omega^2$ (shown by dashed-dotted lines). Another estimate for the same property can be obtained from the rod-climbing measurements by first assuming $\Psi_{2,0}=0$ Pa.s$^2$, which gives $\Psi_{1,0}=2\hat{\beta}$ (shown by dashed lines). These two estimates must match exactly in the absence of a second normal stress difference in the fluid. However, a finite non-zero $\Psi_{2,0}$ is present, as indicated by the significant separation between the two estimated curves. Thus, the rod-climbing rheometer measurements for $\hat{\beta}_{exp}$ in conjunction with the SAOS master curve data can be used to estimate $\Psi_{2,0}=\frac{1}{2}\hat{\beta}_{exp}-\frac{1}{4}\Psi_{1,0}$ as tabulated in Table~\ref{tab:results}. Normal force measurements of $N_1(\dot{\gamma})$ from a 40 mm 2$^\circ$ CP geometry are also shown with filled squares. The normal force measurements cannot access the anticipated second-order scaling of $N_1$ at low shear rates due to limits on the sensitivity of the axial force transducer. }
\end{figure}

This na\"ive \textit{a priori} estimate of $\Psi_{1,0}=2\hat{\beta}_{exp}$ for the various PIB solutions utilized in this study is represented in Fig.~\ref{fig:reconciling} by the dashed lines. If $\Psi_{2,0}$ is indeed zero (as has been observed for some dilute Boger fluids experimentally \cite{keentok1980measurement,magda1991second}), we would expect these dashed lines (i.e., the lower bound prediction of $\Psi_{1,0}$ extracted from rod-climbing experiments) to coincide exactly with the dashed-dotted lines in Fig.~\ref{fig:reconciling}, which depict the actual values of $\Psi_{1,0}$ obtained in Sec.~\ref{sec:SAOS} from the asymptotic quadratic scaling of the SAOS data. However, there is a significant offset in these two \textit{independent} experimental estimates; hence, the initial assumption that $\Psi_{2,0}=0$ is incorrect. A finite non-zero $\Psi_{2,0}$ affects rod-climbing, and in particular, negative values of $\Psi_{2,0}$ increase the value of $\Psi_{1,0}$ that are consistent with a given experimental observation $\hat{\beta}_{exp}$ (see Eq.~\ref{eq:eq7re}). This additional contribution can be obtained by solving Eq.~\ref{eq:eq7} with the independent knowledge of (i) $\Psi_{1,0}$ from the SAOS measurements in Sec.~\ref{sec:SAOS}, and (ii) $\hat{\beta}_{exp}$ from the rod-climbing measurements in Sec.~\ref{sec:RodClimbing}. The values of $\Psi_{2,0}$ thus obtained for various PIB solutions are summarized in Table~\ref{tab:results}. 

As an additional independent check, we also directly measure the material function $N_1(\dot{\gamma})$ using a 40 mm 2$^\circ$ cone-and-plate (CP) geometry, and this data is also presented in Fig.~\ref{fig:reconciling}. As discussed in the Introduction, due to the lower sensitivity limit of $F_{CP,min} \approx 0.001$ N (0.1 gm-force) on the axial thrust measurement, $N_1(\dot{\gamma})$ values $\lesssim 1.6$ Pa.s$^2$ cannot be measured using a 40 mm 2$^{\circ}$ CP geometry. This limit is shown by a horizontal dotted line in Fig.~\ref{fig:reconciling}. Thus, Fig.~\ref{fig:reconciling} pictorially illustrates that the asymptotic second-order regime in which $N_1\left( \dot{\gamma}  \right) \simeq \Psi_{1,0}\dot{\gamma}^2$ cannot be directly accessed using axial force measurements with this CP geometry due to the lower sensitivity limits of the normal force transducer. Larger plates, of course, lower this bound, as is evident by considering Eq.~\ref{eq:eq1}, but the largest available plates ($R_{max} \approx 60$ mm) only result in lowering the dotted line by a factor of 2.25.

From Table~\ref{tab:results}, we observe that the values of $\psi_0$ obtained for all the PIB solutions investigated here lie between the two limiting values 2/7 ($\approx 0.285$) and 1/7 ($\approx 0.143$) predicted by reptation theory \cite{kimura1981polym} (with and without the independent alignment approximation, respectively). Very careful measurements with distributed pressure measurements across a cone and plate have shown that the typical $\psi_0$ values of semi-dilute and concentrated entangled polystyrene solutions are similar and close to the reptation prediction of 2/7 \cite{magda1993rheology}. A value $1/7 < \psi_0 < 2/7$ is typical for semi-dilute polystyrene solutions \cite{magda1993rheology}. 

The exact limiting condition required at low shear rates to observe rod-climbing can be derived analytically by neglecting the inertia term in Eq.~\ref{eq:eq6} and rearranging to show that for rod climbing to be observed, we require\cite{castro1984elementary,lodge1988weissenberg, hu1990climbing} 
\begin{equation}\label{eq:eq12}
   \hat{\beta} > 0 \implies \psi_0 < 0.25.
\end{equation}
Hence, if the independent alignment approximation is exactly obeyed so that $\psi_0 \approx 0.285$, rod-climbing will not be observed and vice-versa. Prima facie, this seems like a serious limitation on the utility of rod-climbing as a technique for measuring $\Psi_{2,0}$. Furthermore, for the least viscoelastic 0.3 wt.\,\% PIB solution, our independent measurements of rod-climbing, the SAOS tTS master curve, and $N_1(\dot{\gamma})$ all suggest $\psi_0=0.205$, a value which satisfies the asymptotic rod-climbing condition in Eq.~\ref{eq:eq12}; however, Fig.~\ref{fig:allheight} reveals that the 0.3 wt.\,\% PIB solution undergoes rod-descending instead of rod-climbing. These experimental observations motivate a more complete understanding of Eq.~\ref{eq:eq6}, and this is considered in the following subsection. 

\subsection{The climbing condition}\label{sec:ClimbingCond}
\begin{figure}
    \centering
    \includegraphics[width=0.47\textwidth]{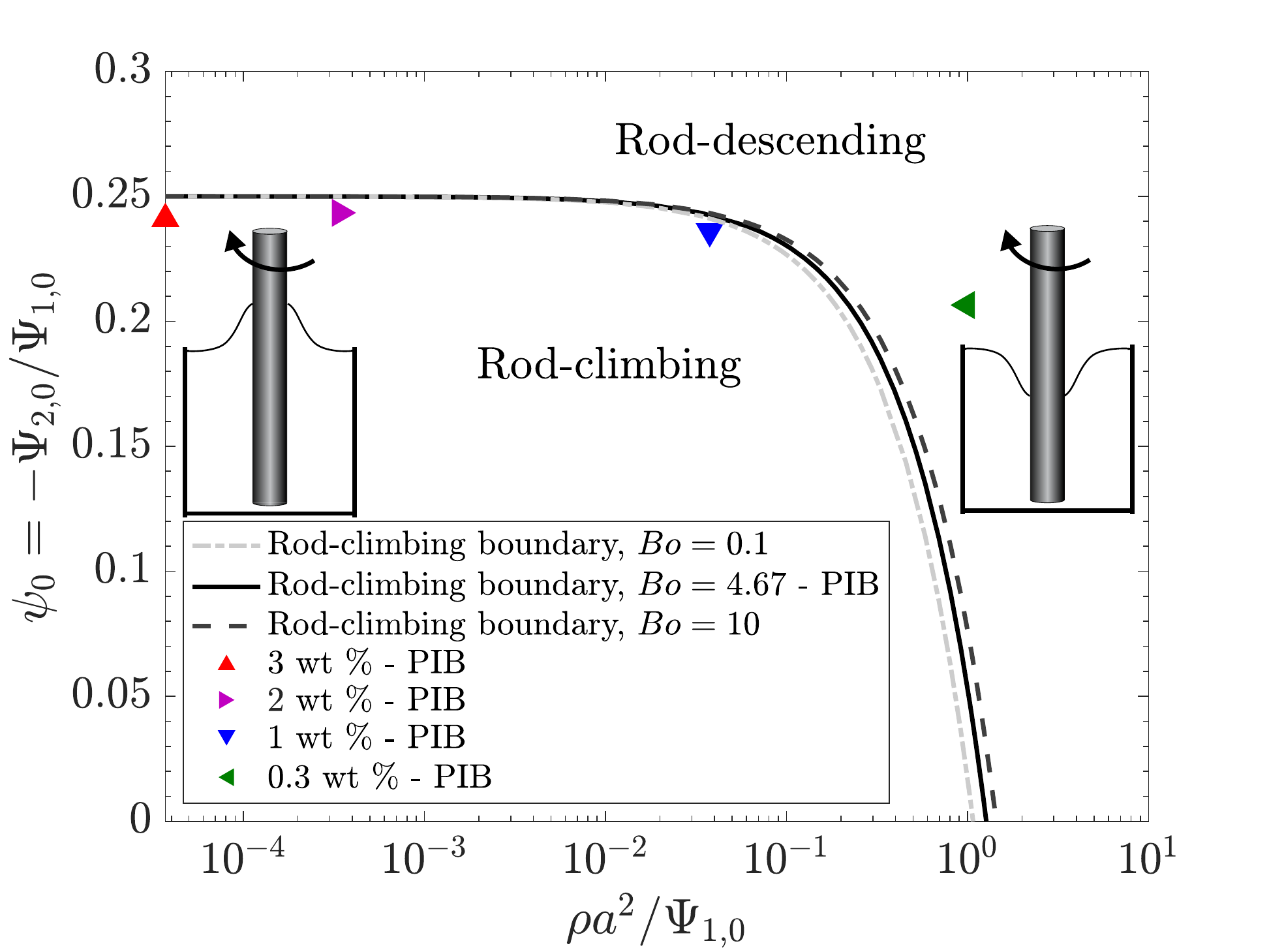}
    \caption{The condition for observing rod-climbing vs. rod-descending is plotted for different Bond numbers, $Bo$, as a competition between the normal stress ratio $\psi_0$ and the dimensionless inertioelastic parameter $\rho{a}^2/\Psi_{1,0}=1/(2El)$ with $El$ being the elasticity number. The curves show the condition of Eq.~\ref{eq:eq13} for three different values of the Bond number $Bo=\rho{g}a^2/\Gamma$ including the value $Bo=4.7$ expected for the PIB solutions used in this study. The boundary line separates rod-climbing from rod-descending. The 3 wt.\,\%, 2 wt.\,\%, and 1 wt.\,\% solutions satisfy the rod-climbing condition and climb the rotating rod, while the 0.3 wt.\,\% solution does not satisfy the rod-climbing condition; hence, its interface descends close to the rotating rod (See Fig.~\ref{fig:allheight}).}
    \label{fig:climbingcond}
\end{figure}

Here we modify the rod-climbing condition in Eq.~\ref{eq:eq12} by accounting for the combined effects of inertia and elasticity. In doing so, we extend the utility of rod-climbing experiments for measuring $\Psi_{2,0}$ to cases where rod-descending is observed. The first step is to return our attention to Eq.~\ref{eq:eq6}, which predicts the small perturbations in the interface height with increasing $\Omega^2$ in the small rotation speed limit. The first term in Eq.~\ref{eq:eq6} given by $4\hat{\beta}/(4+\sqrt{Bo})$ (Term I) suggests that positive perturbations to $h_s(a,\alpha)$ will be observed if the fluid has significant elasticity, i.e., $\hat{\beta}>0$. On the other hand, the second term in Eq.~\ref{eq:eq6} $\rho{a}^2/(2+\sqrt{Bo})$ (Term II) suggests that negative perturbations to $h_s(a,\alpha)$ arising from fluid inertial effects will be observed. As a result, the fluid interface may climb or descend a rotating rod depending on whether Term I $>$ Term II or vice versa. From the climbing condition, $\textrm{d}\Delta{h}(\Omega,a)/\textrm{d}\Omega^2>0$, after rearranging the various terms in Eq.~\ref{eq:eq6} we can obtain the following condition for rod-climbing to be observed:
\begin{equation}\label{eq:eq13}
    \psi_0 < \frac{1}{4}\left( 1-\frac{4+\sqrt{Bo}}{2(2+\sqrt{Bo})}\frac{\rho{a}^2}{\Psi_{1,0}} \right).
\end{equation}
This rod climbing constraint incorporates the competition between inertial and elastic effects, as depicted in Fig.~\ref{fig:climbingcond}. The ratio $\rho{a^2}/\Psi_{1,0}$ represents the relative contributions of fluid inertia and elasticity. Recognizing that for a second order fluid, we can write the relaxation time as $\tau=\Psi_{1,0}/(2\eta_0)$, we can rewrite this ratio in terms of a Deborah number $De=\tau{\Omega}$ and a Reynolds number $Re=\rho\Omega{a}^2/\eta_0$ or alternatively in terms of the elasticity number $El=De/Re=(\tau\Omega)/(\rho\Omega{a}^2/\eta_0)=\Psi_{1,0}/(2\rho{a}^2)$. The curves in Fig.~\ref{fig:climbingcond} show the condition of Eq.~\ref{eq:eq13} for three different values of the Bond number $Bo=\rho{g}a^2/\Gamma$, including the value $Bo=4.7$ appropriate for our PIB solutions. Because of the functional form of the fractional term involving $Bo$ in Eq.~\ref{eq:eq13}, the boundary between rod-climbing and rod-descending is only weakly sensitive to gravitational effects and is predominantly controlled by inertial effects. We also show the actual values of $\psi_0$ and $\rho{a}^2/\Psi_{1,0}$ determined experimentally for the various PIB solutions studied here. The original rod-climbing condition in Eq.~\ref{eq:eq12} is recovered when inertia effects are negligible compared to elasticity effects, i.e., $\rho{a}^2/\Psi_{1,0} \ll 1$.

Fig.~\ref{fig:climbingcond} shows that the values of the material functions determined experimentally for the 3 wt.\,\%, 2 wt.\,\%, and 1 wt.\,\% solutions satisfy the rod-climbing condition and should exhibit positive perturbations to the static interface shape under low finite rod-rotation speeds. This is indeed true as depicted in Fig.~\ref{fig:allheight} by the increase in $\Delta{h}$ with $\Omega^2$ for these three solutions. On the other hand, the low elasticity of the 0.3 wt.\,\% solution satisfies the rod-descending condition as shown in Fig.~\ref{fig:climbingcond} and should exhibit negative perturbations to the static interface shape. This is again found to be true in Fig.~\ref{fig:allheight}, which shows a decrease in $\Delta{h}$ with increasing $\Omega^2$ for the 0.3 wt.\~ \% solution. Also, for the 0.3 wt.\,\% solution, the inertioelastic quantity $\rho{a}^2/\Psi_{1,0}>1$, indicating that inertial effects cannot be ignored. As a result, using the original simplified climbing condition given in Eq.~\ref{eq:eq12} fails to predict the observed rod-descending as it was derived by ignoring inertial effects \cite{lodge1988weissenberg}.  

As indicated in Fig.~\ref{fig:climbingcond}, Eq.~\ref{eq:eq13} predicts whether the small perturbations $\Delta{h}(\Omega, a)$ will be positive or negative in the small $\Omega$ limit. However, it should be noted that even if the criterion of Eq.~\ref{eq:eq13} predicts rod-descending, the rotating rod experiments can still be useful in measuring $\Psi_{2,0}$ as long as the fluid wets the rotating shaft, i.e., the contact angle $\alpha<90^\circ$. The negative free surface perturbations $\Delta{h}(\Omega,a)$ can be readily calculated from a sequence of rod-climbing photographs of the interface shape for a wetting fluid. Once $\Delta{h}(\Omega,a)$ vs. $\Omega^2$ data is available, its slope in the low $\Omega^2$ regime can be equated to Eq.~\ref{eq:eq6} to calculate $\hat{\beta}_{exp}$ irrespective of its sign. This is especially useful in weakly elastic fluids when inertial effects compete with elastic effects, e.g., for the 0.3 wt.\,\% PIB solution, or if the normal stress ratio exceeds $\psi_0>0.25$, e.g., for fluids following the predictions of reptation theory with the independent alignment approximation \cite{magda1993rheology}. The analysis presented in this section and our measurements for the 0.3 wt.\,\% solution, which undergoes rod-descending, extend the validity of rod-climbing rheometry in principle to a much wider range of complex fluids, provided special care is taken in selecting a rigid rod constructed from a solid material that the fluid wets \cite{chandra2022contact}.

Finally, we note that a special (redundant) case arises when $\hat{\beta}=0$, i.e., we can either have $\psi_0=0.25$ or an inelastic fluid with $\Psi_{2,0}=\Psi_{1,0}=0$. In this case, if the fluid is viscoelastic, a finite value of $\Psi_{1,0}$ can first be measured independently as discussed in Sec.~\ref{sec:SAOS} using normal force measurements, or from the asymptotic scaling of a viscoelastic master curve for $G'(\omega_r)$. If no rod-climbing is observed, we can conclude that $\Psi_{2,0} \approx -0.25\Psi_{1,0}$ in such a fluid. The interface shape will be unperturbed at low rod rotation speeds provided $\rho{a}^2/\Psi_{1,0} \ll 1$ (i.e., weak inertial effects). If the inertial effects are strong, i.e., $\rho{a}^2/\Psi_{1,0}>1$, then this fluid will exhibit rod-descending at all rod rotation speeds. 

If the independent measurement of the first normal stress difference reveals that the fluid is essentially inelastic, i.e., $\Psi_{1,0} \approx 0$, then a fluid with $\hat{\beta}=0$ will simply exhibit Newtonian rod-descending at sufficiently high rotation rates as shown by the dashed pink line in Fig.~\ref{fig:allheight}. In this case, we can conclude that $\Psi_{2,0} \simeq \Psi_{1,0}=0$.



The observations discussed in this section imply that calling this technique ``Rod-climbing rheometry'' might be misleading as empirical measurements are still useful even when the fluid might experience rod-descending. Hence, a more general description, ``rotating rod rheometry,'' is more suitable than rod-climbing rheometry, as the protocol presented in this study is readily applied irrespective of whether a fluid undergoes rod-climbing or rod-descending.

\subsection{Varying fluid elasticity and viscosity}\label{sec:Boger}

\begin{figure}
    \centering
    \includegraphics[width=0.47\textwidth]{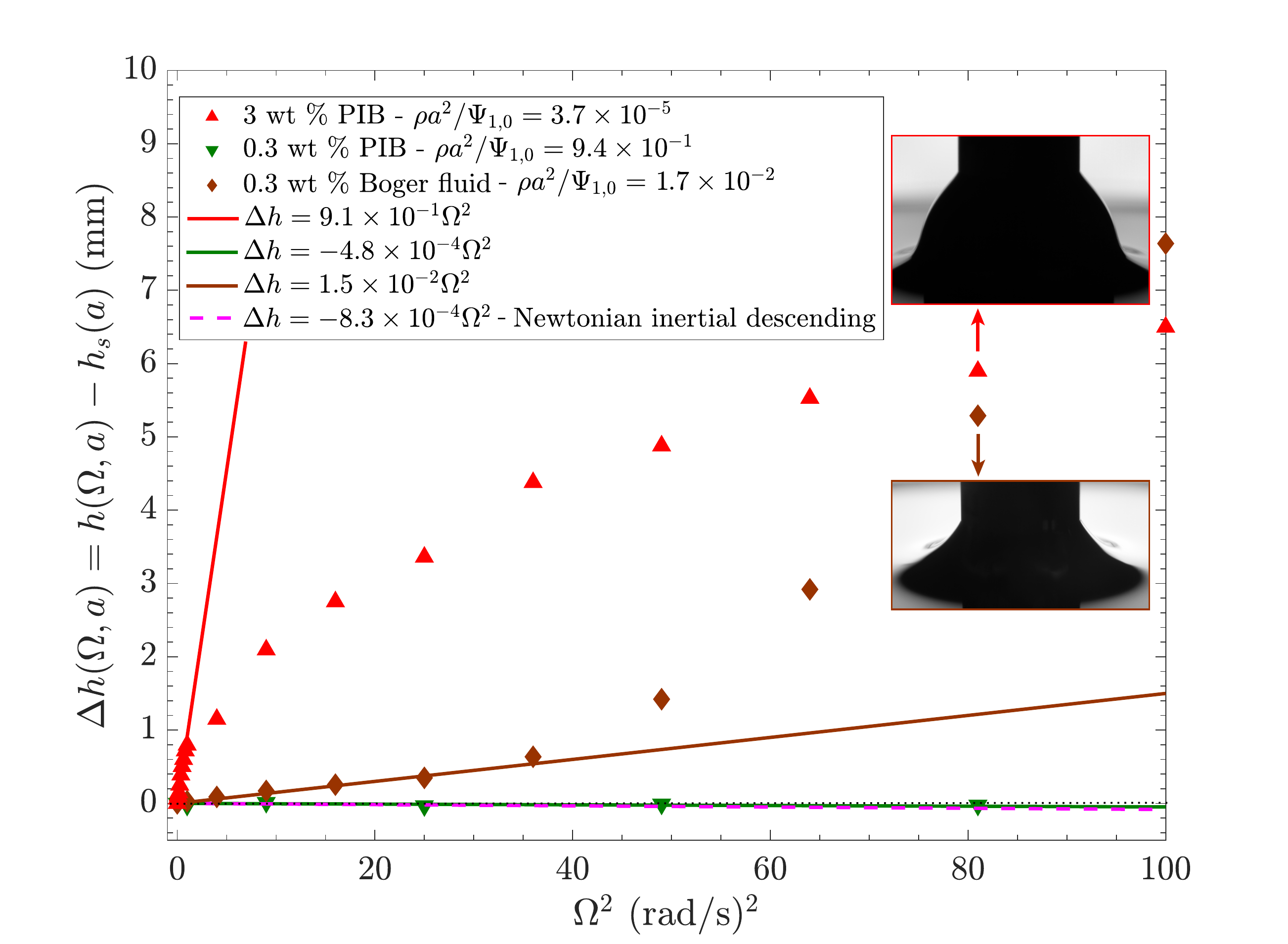}
    \caption{Change in the interface height at the rotating rod ($\Delta{h}(\Omega,a)$) compared to the static rise $h_s(a)$ for the 3.0 and 0.3 wt.\,\% PIB solutions utilized in this study compared with a 0.3 wt.\,\% Boger fluid. Increasing the fluid elasticity dramatically enhances rod climbing in the 0.3 wt.\,\% Boger fluid. The weakly elastic 0.3 wt.\,\% PIB solution does not satisfy the condition for climbing (see Fig.~\ref{fig:climbingcond}), and hence, undergoes rod-descending. The dashed line depicts the rod-descending exhibited by a purely Newtonian fluid without any elasticity, which arises solely due to the inertial effects and hence is the lower bound for the $\Delta{h}$ vs. $\Omega^2$ curves. Measurements of $\Delta{h}$ vs. $\Omega^2$ curves for a fluid with finite non-zero normal stresses will lie above this lower bound, although the difference is small on the scale shown here (cf. Fig.~\ref{fig:allheight}). Insets show the interface shapes for 3 wt.\,\% PIB and 0.3 wt.\,\% Boger fluid at $\Omega=9$ rad/s.}
    \label{fig:Bogerheight}
\end{figure}

To further corroborate the accuracy of the modified climbing condition presented in Eq.~\ref{eq:eq13}
and support the ideas discussed in the previous subsection, we artificially modify the elasticity of the 0.3 wt.\,\% PIB solution to change the relative balance of elastic and inertial effects in the rotating rod experiment. From bead-spring theory for dilute solutions, we know that the elasticity in a polymeric solution scales as $\Psi_{1,0} \simeq 2\eta_P\tau_s$, where $\eta_P$ is the polymer contribution to the viscosity, and $\tau_s$ is the shear relaxation time. In a dilute solution, we expect $\Psi_{2,0}=0$. The magnitude of viscoelastic effects can be varied by either 1) increasing the polymer concentration or 2) increasing the relaxation time of the fluid. In the former case, we anticipate that $\eta_P \sim (\textrm{C}/\textrm{C}^*)^{2.4/(3\nu-1)}$ for a semi-dilute entangled solution in a good solvent with $\nu = 0.588$ \cite{heo2005scaling} but the magnitude of second normal stress difference coefficient will also increase, and the expected functional form of $\psi_0$ is not known. However, increasing the shear relaxation time $\tau_s$ and remaining in the dilute regime can also be achieved by increasing the solvent viscosity $\eta_s$. According to Rouse-Zimm bead spring theories, the relaxation time will increase linearly with $\eta_s$ \cite{larson2013constitutive}. We have utilized technique 1 to vary the elasticity in the PIB solutions so far in this study. It also explains the weak elasticity in the semi-dilute 0.3 wt.\,\% PIB solution compared to more concentrated ones. Technique 2 is the standard recipe for preparing Boger fluids \cite{boger1977highly} and has been widely used to prepare highly elastic fluids at C $\lesssim$ C$^{*}$ with a constant viscosity \cite{prilutski1983model,magda1991second}. Hence, to augment the elasticity of the 0.3 wt.\,\% PIB solution, we increase the solvent viscosity $\eta_s$ by mixing a viscous polyalphaolefin oil (PAO) with the paraffin solvent, which increases $\eta_s$ and consequently, $\tau_Z$ and $\Psi_{1,0}$, by around two orders of magnitude. We identify this formulation by the label 0.3 wt.\,\% PIB Boger fluid as it has the same concentration of PIB as the weakly elastic shear thinning 0.3 wt.\,\% PIB solution studied in Sec.~\ref{sec:SAOS}$-$\ref{sec:Reconcile} but a higher viscosity. 


\begin{table}[t!]
\small
  \caption{\ Measurements of the climbing constant $\hat{\beta}$ from the rod-climbing rheometry and $\Psi_{2,0}$ by combining rod-climbing rheometry with the SAOS measurements from a conventional rheometer (TA instruments ARES-G2) for the 0.3 wt.\,\% PIB Boger fluid (PIB in polyalphaolefin and paraffinic oil) used in this study. Results from two previous studies\cite{hu1990climbing,magda1991second} for a similar PIB Boger fluid (PIB in polybutene and 2-chloropropane) but with different concentrations are also tabulated for comparison.}
  \label{tab:resultsBoger}
  \begin{tabular*}{0.48\textwidth}{@{\extracolsep{\fill}}l c c c c}
   \hline
    C (wt.\,\%) & $\hat{\beta}$ (Pa.s$^2$) & $\psi_0=-\frac{\Psi_{2,0}}{\Psi_{1,0}}$ & $\Psi_{1,0}$ (Pa.s$^2$) & $\Psi_{2,0}$ (Pa.s$^2$) \\
  \hline
    0.30 (This study) & 0.27 & 0.13 & 1.16 & $-0.16$ \\
    0.24\cite{hu1990climbing} & 1.68 & 0.11 & 6.00 & $-0.66$ \\
    0.10\cite{magda1991second} & 1.28 & 0.01 & 2.65 & $-0.03$ \\
   \hline
  \end{tabular*}
\end{table}

Fig.~\ref{fig:Bogerheight} shows the dramatic effect of increasing the fluid elasticity on its rod-climbing ability. For comparison, the results for the highly elastic 3 wt.\,\% PIB and weakly elastic 0.3 wt.\,\% PIB solutions are also presented here. The weakly elastic 0.3 wt.\,\% PIB solution, which undergoes rod-descending due to the dominance of inertial effects, now becomes strongly rod-climbing when it is ``bogerized'' (i.e., converted to a Boger fluid) solely by increasing the solvent viscosity while retaining the same PIB concentration of 0.3 wt.\,\%. In fact, at a rotation rate of 10 rad/s, the 0.3 wt.\,\% Boger fluid now climbs the rod to a greater height than the 3 wt.\,\% semi-dilute entangled fluid. This is because of the large value of $\Psi_{1,0}$, but the very small value of $\Psi_{2,0}$ in this dilute solution. We can follow the rotating rod rheometry protocol established in Sec.~\ref{sec:Results} to calculate $\Psi_{1,0}$ and $\Psi_{2,0}$ in the 0.3 wt.\,\% Boger fluid. The results are tabulated in Table~\ref{tab:resultsBoger}, and $\hat{\beta}$ increases from a value of $\hat{\beta}_{exp}\simeq 2 \times 10^{-3}$ Pa.s$^{2}$ for the 0.3 wt.\,\% PIB solutions to a value $\hat{\beta}_{exp}\simeq 0.27$ Pa.s$^{2}$. Also shown in the table just for comparison are results from two previous rod-climbing measurements \cite{hu1990climbing,magda1991second} of similar PIB-based (PIB molecular weight $\approx 10^6$ g/mol.) Boger fluids at slightly lower concentrations. 

Applying the rod-climbing condition Eq.~\ref{eq:eq13} to this 0.3 wt.\,\% Boger fluid reveals that it lies deep in the rod-climbing region in Fig.~\ref{fig:climbingcond} with $\psi_0 = 0.13$ and $\rho{a}^2/\Psi_{1,0}=1.55 \times 10^{-2}$, which rationalizes the dramatic  transition from rod-descending to rod-climbing that can be engineered into this 0.3 wt.\,\% polymer solution simply by the addition of a viscous solvent. 

\section{Conclusions}
We have revisited the rod-climbing rheometer\cite{beavers1975rotating} for measuring normal stress differences in complex fluids that was originally proposed around four decades ago \cite{joseph1973free2}. In doing so, we integrate its performance with modern-day torsional rheometers to facilitate self-consistent predictions of the zero shear rate values of both the first and the second normal stress coefficients $\Psi_{1,0}$ and $\Psi_{2,0}$, which are often very challenging to determine accurately. The protocol for rotating rod rheometry presented here involves:
\begin{enumerate}
    \item Evaluating the first normal stress difference coefficient in the limit of zero shear rate ($\Psi_{1,0}$) from SAOS master curve data by using the asymptotic result from simple fluid theory $\Psi_{1,0} = \lim_{\omega\to0}2G'/\omega^2$.

    \item Determining the climbing constant $\hat{\beta}_{exp}$ of the fluid by measuring the rate of change of perturbations to the static interface $\Delta{h}(\Omega,a)$ vs. $\Omega^2$, i.e., ($\textrm{d}\Delta{h}/\textrm{d}\Omega^2$)$_{exp}$ for small $\Omega$. 

    \item By equating the value of $\hat{\beta}_{exp}$ determined experimentally to the theoretical result (Eq.~\ref{eq:eq6}) of the second order fluid theory, the second normal stress coefficient can then be determined from the two independent measurements using the relationship $\Psi_{2,0} = \frac{1}{2}\hat{\beta}_{exp}-\frac{1}{4}\Psi_{1,0}$. 
\end{enumerate}

We have used this protocol to determine $\Psi_{1,0}$ and $\Psi_{2,0}$ of several PIB solutions in the concentrated and semi-dilute regimes. We observe $\psi_0 = -\Psi_{2,0}/\Psi_{1,0}<0.25$ for all the PIB solutions, so all of them should exhibit rod-climbing according to the original rod-climbing criterion obtained by neglecting inertial effects \cite{lodge1988weissenberg} (Eq.~\ref{eq:eq12}). We indeed observe rod-climbing for the more concentrated solutions; however, the least concentrated 0.3 wt.\,\% weakly elastic PIB solution exhibited rod-descending, hinting at substantial inertial effects. Hence, we have modified the rod-climbing condition by considering the relative strength of inertial effects compared to the elastic effects, as quantified by the ratio $\rho{a}^2/\Psi_{1,0}$ (Eq.~\ref{eq:eq13}). The modified rod-climbing condition successfully rationalizes the observed dipping of the free surface for the 0.3 wt.\,\% weakly viscoelastic fluid.

To elucidate the competition between the elastic and inertial effects in determining whether a given fluid will exhibit rod-climbing or rod-descending, we deliberately enhanced the elasticity of the 0.3 wt.\,\% PIB solution by ``bogerizing'' it through the addition of a more viscous solvent, i.e., we prepared a 0.3 wt.\,\% PIB Boger fluid. The resulting highly elastic fluid has a higher value of $\Psi_{1,0}$ and a lower value of $\psi_0$. It, therefore, undergoes pronounced rod-climbing due to the dominant effect of elasticity overwhelming inertial effects, in marked contrast to the weakly elastic 0.3 wt.\,\% PIB solution. 

Thus, we conclude that if a fluid undergoes rod-descending instead of rod-climbing, it does not conclusively indicate an absence of elastic effects in a fluid. Weak elasticity might still be present but is largely masked by the dominance of inertial effects, resulting in the quadratic, but negative, variation in the free surface height we observed with the 0.3 wt.\% PIB fluid. Even in such a case, the weak contributions of $\Psi_{1,0}$ and $\Psi_{2,0}$ can still be extracted using the protocol presented in this study from the negative slope $\textrm{d}\Delta{h}/\textrm{d}\Omega^2$.

In conjunction with time-Temperature Superposition (tTS) measurements of a linear viscoelastic master curve of $G'(\omega_r)$, our analysis and results show that the rotating rod experiment can be very useful in extending measurements of the normal stress differences in complex fluids to lower shear rate limits, irrespective of whether they result in rod-climbing or rod-descending, by allowing for the inertial contributions to the interface shape and ensuring the rod material is selected such that the fluid is wetting. Hence, a general description, ``rotating rod rheometry'' for this technique, is more apt.

\section*{Appendix A: Climbing height of a second-order fluid on a slowly rotating thin rod}\label{sec:AppA}

\begin{figure}[h]
    \centering
    \includegraphics[width=0.4\textwidth]{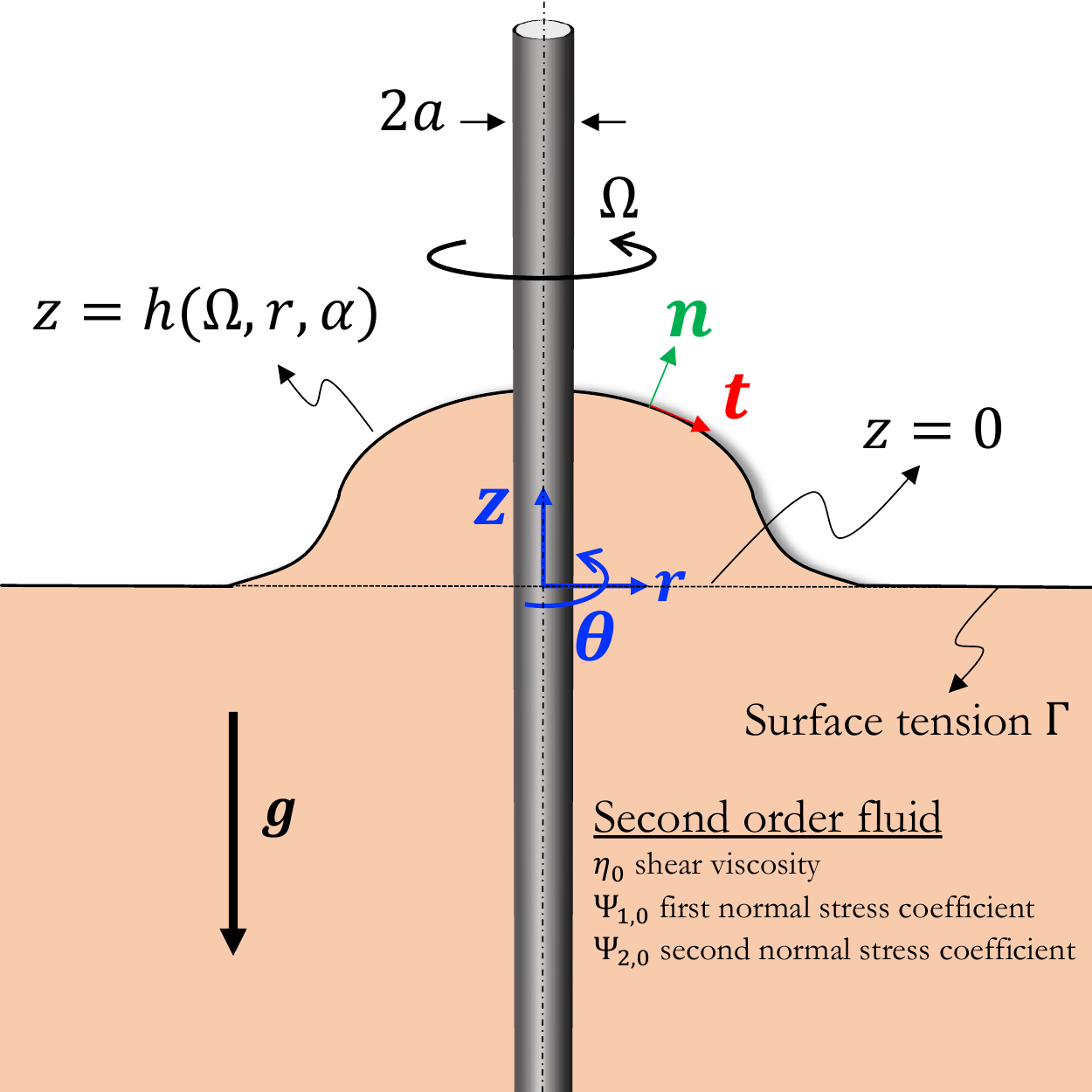}
    \caption{Schematic of the rotating rod problem in a cylindrical coordinate system $(r,\theta,z)$ with $(\mathbf{e}_r,\mathbf{e}_{\theta},\mathbf{e}_z)$ as the unit vectors in the respective directions. A thin rod of radius $a$ with its axis along the z-axis rotates around its axis with a constant rotational velocity $\Omega$. $\mathbf{n}$ and $\mathbf{t}$ denote the unit vectors normal and tangent to the interface $z=h(\Omega,r,\alpha)$.}
    \label{fig:derivation}
\end{figure}

Using modern notation and retaining inertial contributions, this appendix reproduces a formal derivation of Eq.~\ref{eq:eq3} using a domain perturbation analysis method from the original works of D. D. Joseph and co-workers \cite{joseph1973free1,joseph1973free2}. The problem setup is depicted in Fig.~\ref{fig:derivation}. A thin rod of radius $a$ is submerged in a semi-infinite pool of a second-order incompressible fluid with density $\rho$, surface tension $\Gamma$, and contact angle $\alpha$ with the rod. The rod is infinitely long and rotates with a constant angular velocity $\Omega$. The fluid surface is exposed to atmospheric pressure $p_a$ and deviates from its initial static shape $z=h_s(r,\alpha)$ to a steady profile $z=h(\Omega,r,\alpha)$ due to the shear flow generated by the rod rotation. The profile $z=h(\Omega,r,\alpha)$ is determined by the combined action of normal stresses, inertia, surface tension, and gravity. For an axially symmetric velocity $\mathbf{u}=v(r,z)\mathbf{e}_{\theta}+\mathbf{\tilde{{u}}}$ with $\mathbf{\tilde{{u}}}=u(r,z)\mathbf{e}_r+w(r,z)\mathbf{e}_z$ in the cylindrical coordinates $[r,\theta,z]$, the continuity and Navier-Stokes equations with ${\sigma}_{ij}$ as the stress tensor for a second-order fluid can be written as: 

\begin{subequations}\label{eq:eq14}
\begin{equation}\label{eq:eq14a}
    {\partial_r({ru})} + r{\partial_z({w})} = 0, 
\end{equation} 
\begin{equation}\label{eq:eq14b}
    \rho \left[ u\partial_r{u} + w\partial_z{u} -\frac{u^2}{r} \right] = -\partial_r\Phi + \partial_r{\sigma_{rr}}+\partial_z{\sigma_{rz}}+\frac{1}{r}(\sigma_{rr}-\sigma_{\theta\theta}), 
\end{equation} 
\begin{equation}\label{eq:eq14c}
        \rho \left[ u\partial_r{v} + w\partial_z{v} +\frac{uv}{r} \right] = \frac{1}{r^2}\partial_r(r^2\sigma_{r\theta})+\partial_z\sigma_{z\theta},
\end{equation}
\begin{equation}\label{eq:eq14d}
    \rho[u\partial_r{w}+w\partial_z{w}] = -\partial_z\Phi + \partial_r\sigma_{rz}+\partial_z\sigma_{zz}+\frac{1}{r}\sigma_{rz},
\end{equation}
where the stress tensor is given by 
\begin{equation}\label{eq:eq14e}
    \boldsymbol{\sigma} = \eta_0 \mathbf{A}_1 - 0.5 \Psi_{1,0} \mathbf{A}_2 + (\Psi_{1,0}+\Psi_{2,0}) \mathbf{A}_1^2.
\end{equation}
\end{subequations}
Here $\Phi=p+\rho{g}z$ is the pressure head, $\partial_i = \partial/\partial{x}_{i}$. $\mathbf{A}_1(\mathbf{u})=(\boldsymbol \nabla \mathbf{u} + \boldsymbol \nabla \mathbf{u}^T)$, and $\mathbf{A}_2(\mathbf{u})=(\mathbf{u}\cdot \boldsymbol \nabla)\mathbf{A}_1 + \mathbf{A}_1\boldsymbol \cdot \nabla\mathbf{u} + \boldsymbol \nabla\mathbf{u} ^T \cdot \mathbf{A}_1$ are the first two Rivlin–Ericksen tensors. The coefficients $\eta_0$, $\Psi_{1,0}$, and $\Psi_{2,0}$ are the viscosity, first and second normal stress difference coefficients of the fluid in the limit of zero shear. The unit normal to the free interface $z=h(\Omega,r,\alpha)$ is given by $\mathbf{n}=\frac{-h'}{\sqrt{1+h'^2}}\mathbf{e}_r+\frac{1}{\sqrt{1+h'^2}}\mathbf{e}_z$ where $h'=\textrm{d}h/\textrm{d}r$. The two orthogonal tangential vectors to the free interface are $\mathbf{e}_{\theta}$ and $\mathbf{t}=-\frac{1}{\sqrt{1+h'^2}}\mathbf{e}_r-\frac{h'}{\sqrt{1+h'^2}}\mathbf{e}_z$. The solution to Eq.~\ref{eq:eq14} must satisfy the following boundary conditions:

\begin{subequations}\label{eq:eq15}
\textrm{No slip at the rod:}
    \begin{equation}\label{eq:eq15a}
        \mathbf{u} = a\Omega\mathbf{e}_{\theta} \textrm{ at } r=a.
    \end{equation}

\textrm{No flux normal to the interface:}
   \begin{equation}\label{eq:eq15b}
        w-uh'=0 \textrm{ at } z=h(\Omega,r,\alpha).
    \end{equation}

\textrm{No tangential stress at the fluid interface:}
   \begin{equation}\label{eq:eq15c}
       \sigma_{n\theta} = \sigma_{z\theta} - h'\sigma_{r\theta} = 0 \textrm{ at } z=h(\Omega,r,\alpha), \textrm{ and}
    \end{equation}
   \begin{equation}\label{eq:eq15d}
        \sigma_{nt} = h'(\sigma_{zz}-\sigma_{rr}) + (1-h'^2)\sigma_{rz} = 0 \textrm{ at } z=h(\Omega,r,\alpha).
    \end{equation}

\textrm{The normal stress jump at the interface is balanced by the surface tension ($\Gamma$) force:}
\begin{equation}\label{eq:eq15e}
    p_a-\Phi+\sigma_{zz}-h'\sigma_{rz} + \rho{g}h = \frac{\Gamma}{r} \left[ \frac{rh'}{\sqrt{1+h'^2}} \right]' \textrm{ at } z=h(\Omega,r,\alpha).
\end{equation}

\textrm{Contact angle condition:}
\begin{equation}\label{eq:eq15f}
    h'(\Omega,r,\alpha) = \textrm{cot}(\alpha)  \textrm{ at } r=a.
\end{equation}

\textrm{Finally, the solution approaches the hydrostatic solution with a flat free interface as $r \to \infty$:}
\begin{equation}\label{eq:eq15g}
    h(\Omega,r,\alpha) \to 0 \textrm{ as } \mathbf{u} \to 0 \textrm{, and } \Phi\to0.
\end{equation}
\end{subequations}
The free surface problem described by Eq.~\ref{eq:eq14} and~\ref{eq:eq15} can be solved using the domain perturbation method under the condition that the total fluid domain volume is conserved, i.e., $\int_{r=a}^{r\to\infty}rh(\Omega,r,\alpha)\textrm{d}r=0$. The solution then can be expanded as a power series: 
\begin{equation}\label{eq:eq16}
   \begin{bmatrix}
       \mathbf{u} \\
       \boldsymbol{\sigma} \\
       \Phi \\
       h
   \end{bmatrix} = 
   \sum_i   \begin{bmatrix}
       \mathbf{u}^{(i)} \\
       \boldsymbol{\sigma}^{(i)} \\
       \Phi^{(i)} \\
       h^{(i)}
   \end{bmatrix}\Omega^i.
\end{equation}

Substituting Eq.~\ref{eq:eq16} in the governing equations Eq.~\ref{eq:eq14}, we get the following zero-th order governing equations:
\begin{subequations}\label{eq:eq17}
   \begin{equation}\label{eq:eq17a}
    (\mathbf{u}^{(0)} \cdot \boldsymbol \nabla) \mathbf{u}^{(0)} = -\mathbf{\nabla} \Phi^{(0)} +  \boldsymbol{\nabla} \cdot \boldsymbol\sigma^{(0)}, 
    \end{equation}
    \begin{equation}\label{eq:eq17b}
        \boldsymbol{\nabla} \cdot \mathbf{u}^{(0)} = 0. 
    \end{equation}
\textrm{Solution to Eq.~\ref{eq:eq17a},~\ref{eq:eq17b} along with the zero-th order boundary conditions Eq.~\ref{eq:eq15} is given by $\mathbf{u}^{(0)}=\mathbf{0}$, $\boldsymbol{\sigma}^{(0)}=\mathbf{0}$, $\Phi^{(0)}=p_a$ and recovers the static interface rise $h^{(0)}=h_s(r,\alpha)$ which can be computed by numerically solving: }
\begin{equation}\label{eq:eq17c}
\frac{\Gamma}{r} \left[ \frac{rh'_s}{\sqrt{1+h_s'^{2}}} \right]' = \rho{g}h_s
\end{equation}
\textrm{subjected to $h'_s(r=a)=\textrm{cot}(\alpha)$ and $h_s\to0$ as $r\to\infty$.}
\end{subequations}

The zero-th order solution obtained by solving Eq.~\ref{eq:eq17} can be used to solve the following first-order governing equations:
\begin{subequations}\label{eq:eq18}
   \begin{equation}\label{eq:eq18a}
    \mathbf{u}^{(0)} \cdot \boldsymbol \nabla \mathbf{u}^{(1)} + \mathbf{u}^{(1)} \cdot \boldsymbol \nabla \mathbf{u}^{(0)} = -\mathbf{\nabla} \Phi^{(1)} + \boldsymbol \nabla \cdot \boldsymbol\sigma^{(1)}, 
    \end{equation}
    \begin{equation}\label{eq:eq18b}
        \boldsymbol{\nabla} \cdot \mathbf{u}_1 = 0. 
    \end{equation}
\textrm{Solution to Eq.~\ref{eq:eq18a},~\ref{eq:eq18b} along with the first order boundary conditions Eq.~\ref{eq:eq15}a-g is given by $\mathbf{u}^{(1)}=\frac{a^2}{r}\mathbf{e}_{\theta}$, $\boldsymbol{\sigma}^{(1)}=\eta_0(\boldsymbol{\nabla}\mathbf{u}^{(1)}+\boldsymbol{\nabla}\mathbf{u}^{{(1)}^T})=-\eta_0{a}^2/r^2(\mathbf{e}_r\mathbf{e}_{\theta}+\mathbf{e}_{\theta}\mathbf{e}_{r})$, and $\Phi^{(1)}=0$. $h^{(1)}=0$ since $h$ is an even function of $\Omega$, and can be computed by solving: }
\begin{equation}\label{eq:eq18c}
    {\Gamma} \left( {rh{^{(1)}}^{'}} \right)^{'} = \rho{rg}h^{(1)}
\end{equation}
\textrm{subjected to $h{^{(1)}}'(r=a)=0$ and $h^{(1)}\to0$ as $r\to\infty$.}
\end{subequations}

The zero-th and first-order solutions obtained by solving Eq.~\ref{eq:eq17} and~\ref{eq:eq18} can now be used to solve for the following second-order governing equations, which give the very first non-trivial contribution to the interface shape and pave the path to arrive at the final results used in Eq.~\ref{eq:eq3}:
\begin{subequations}\label{eq:eq19}
    \begin{equation}\label{eq:eq19a}
        \rho\mathbf{u}^{(1)} \cdot \boldsymbol{\nabla} \mathbf{u}^{(1)} = \boldsymbol{\nabla} \Phi^{(2)} + \boldsymbol{\nabla} \cdot \boldsymbol{\sigma}^{(2)}
    \end{equation}
    \textrm{where $\boldsymbol{\sigma}^{(2)}=\eta_0\mathbf{A}_1(\mathbf{u}^{(2)})-0.5\Psi_{1,0}\mathbf{A}_2(\mathbf{u}^{(1)})+(\Psi_{1,0}+\Psi_{2,0})[\mathbf{A}_2(\mathbf{u}^{(1)})]^2$ giving:}
    \begin{equation}\label{eq:eq19b}
    \nabla \cdot \boldsymbol{\sigma}^{(2)}=\eta_0\boldsymbol{\nabla}^2\mathbf{u}^{(2)}-(0.5\Psi_{1,0}+2\Psi_{2,0})\frac{8a^4}{r^5}\mathbf{e}_r  =\eta_0\boldsymbol{\nabla}^2\mathbf{u}^{(2)}-\hat{\beta}\frac{8a^4}{r^5}\mathbf{e}_r 
    \end{equation}
    \textrm{Thus, the climbing constant $\hat{\beta}=0.5\Psi_{1,0}+2\Psi_{2,0}$ arises naturally at second order. The solution to Eq.~\ref{eq:eq19a}, and~\ref{eq:eq19a} is given by $\mathbf{u}^{(2)}=\mathbf{0}$, $\boldsymbol{\sigma}^{(2)}=\frac{4a^4}{r^4}\left[ \Psi_{2,0}\mathbf{e}_r\mathbf{e}_r + (\Psi_{1,0}+\Psi_{2,0})\mathbf{e}_{\theta}\mathbf{e}_{\theta}  \right]$, and $\Phi^{(2)}=\frac{2{a}^4}{r^4}\hat{\beta}-\frac{\rho{a}^4}{2r^2}$. Finally, the normal stress jump balance Eq.~\ref{eq:eq15e} at second-order gives:}
    \begin{equation}\label{eq:eq19c}
        \frac{\Gamma}{r}\left( rh{^{(2)}}' \right)' - \rho{g}h^{(2)} = -\frac{2a^4}{r^4}\hat{\beta} + \frac{\rho{a}^4}{2r^2}
    \end{equation}
    \textrm{subjected to $h{^{(2)}}'(r=a)=0$ and $h^{(2)}(r)\to 0$ as $r\to \infty$. To solve Eq.~\ref{eq:eq19c} analytically, it can be rearranged as:}
    \begin{equation}\label{eq:eq19d}
        \mathcal{L}(H) = \mathcal{M}(H)+\Phi^{(2)},
    \end{equation}
    \textrm{where $H=h^{(2)}/a^4$, and the operands $\mathcal{L}(\circ)=r[r(\circ)]'-Bo(\circ)$ with the Bond number $Bo=\rho{a}^2g/\Gamma$, and $\mathcal{M}(\circ)=(r-a^2/r)[r(\circ)']'$. The boundary conditions can be written as $H'(r=a)=0$, $(H,H')\to(0,0)$ as $r\to\infty$. Eq.~\ref{eq:eq19d} can be solved analytically by successive approximations:}
    \begin{equation}\label{eq:eq19e}
    \begin{matrix}
        H = H_0+H_1+H_2+..., \\
        \mathcal{L}H_0=\Phi^{(2)}, \\
        \mathcal{L}H_{n+1}=\mathcal{M}H_n \textrm{     (n=0,1,2)}.
        \end{matrix}
    \end{equation}
    The approximating functions $H_n$ also satisfy the above boundary conditions for $H$. Finally, using the identity $\mathcal{L}(r^{-\beta})=(\beta^2-Bo)r^{-\beta}$ one can find the following solutions: 
    \begin{multline}\label{eq:eq19f}
        H_0(r)=\frac{4a^2\hat{\beta}}{(16-Bo)\Gamma}\left[ \frac{4a^{\sqrt{Bo}-4}}{\sqrt{Bo}\,r^{\sqrt{Bo}}}-\frac{1}{r^4} \right]+ \\ 
        \frac{\rho{a}^2}{(4-Bo)\Gamma}\left[  \frac{1}{r^2} -\frac{2a^{\sqrt{Bo}-2}}{\sqrt{Bo}\,r^{\sqrt{Bo}}} \right], \textrm{ and}
    \end{multline}
    \begin{equation}\label{eq:eq19g}
        H_1(r) = \sum_{i=1}^{5} g_i(r)
    \end{equation}
    where 
    \begin{align}
        g_1 = \frac{4c_1}{4-Bo}\left[ \frac{1}{r^2} - \frac{2a^{\sqrt{Bo}-2}}{\sqrt{Bo}\,r^{\sqrt{Bo}}} \right], \\
        g_2 = \frac{16c_2-4a^2c_1}{16-Bo}\left[ \frac{1}{r^4} - \frac{4a^{\sqrt{Bo}-4}}{\sqrt{Bo}\,r^{\sqrt{Bo}}} \right], \\
        g_3 = -\frac{16a^2c_2}{36-Bo}\left[ \frac{1}{r^6} - \frac{6a^{\sqrt{Bo}-6}}{\sqrt{Bo}\,r^{\sqrt{Bo}}} \right], \\
        g_4 = - \frac{\sqrt{Bo}\,c_3}{2}\,r^{\sqrt{Bo}}\left[ \textrm{ln }(r/a)+1/\sqrt{Bo} \right], \\
        g_5 = -\frac{Bo\,a^2c_3}{4(\sqrt{Bo}+1)r^{\sqrt{Bo}}}\left[ \frac{1}{r^2} - \frac{\sqrt{Bo}+2}{\sqrt{Bo}\,a^2} \right], \\
        c_1 = \frac{\rho{a}^2}{(4-Bo)\Gamma}, 
        \end{align}
        \begin{align}
        c_2 = -\frac{4a^2\hat{\beta}}{(16-Bo)\Gamma}, \\
        \textrm{and } c_3 = -4a^{\sqrt{Bo}-4}c_2/\sqrt{Bo}-2a^{\sqrt{Bo}-2}c_1/\sqrt{Bo}.
    \end{align}
\end{subequations}
Thus, we can find $h^{(2)} \simeq a^4(H_0+H_1+...)$. 

Solutions to the third and fourth-order problems can also be obtained but will not be pursued here. In the third order, a correction to the azimuthal velocity component arises, which does not contribute to the free surface shape alteration as $h$ is an even function of $\Omega$. At the fourth order, the fluid motion departs from the simple Couette type $\mathbf{u}=a\Omega\mathbf{e}_{\theta}$, and velocity corrections in both the axial and radial directions come into the picture. Thus, a discernible secondary flow in the ($r,z$) plane is observed. The free surface profile is also altered at fourth order. The solutions at third and fourth order depend on additional material constants beyond the three already involved till the second-order problem, which introduces additional unknowns to be determined. Hence, we stop our analysis at the second order and derive Eq.~\ref{eq:eq3} from the results obtained so far. 

In summary, in the small $\Omega$ limit, we can approximate the interface shape as: 

\begin{equation}
\begin{aligned}\label{eq:eq20}
    h(\Omega,r,\alpha) &= h^{(0)} + h^{(1)}\Omega + h^{(2)}\Omega^2 + O(\Omega^3) \\
      &= h_s(\Omega,\alpha) + h^{(2)}(\Omega,r)\Omega^2 + O(\Omega^2\alpha+\Omega^4) + ...\\
         &= h_s(\Omega,\alpha) + \left[ \frac{4a^2\hat{\beta}}{(16-Bo)\Gamma} \left( \frac{4a^{\sqrt{Bo}}}{\sqrt{Bo}\,r^{\sqrt{Bo}}} - \frac{a^4}{r^4} \right) \right. \\
  & \left. + \frac{\rho{a}^4}{2(4-Bo)\Gamma}\left( \frac{a}{r^2} - \frac{2a^{\sqrt{Bo}}}{\sqrt{Bo}\,r^{\sqrt{Bo}}} \right) \right] {\Omega^2} + O(\Omega^2\alpha+\Omega^4) + ...,
\end{aligned}
\end{equation}
which at $r=a$ gives Eq.~\ref{eq:eq3}:
\begin{equation}\label{eq:eq21}
\begin{aligned}
    h(\Omega,a,\alpha) = h_s(a,\alpha) + \frac{a}{2\left( \Gamma\rho{g} \right)^{1/2}}\left[ \frac{4\hat{\beta}}{4+\sqrt{Bo}}-\frac{\rho{a}^2}{2+\sqrt{Bo}} \right]\Omega^2 \\
    + O(\Omega^2\alpha+\Omega^4) + ...
    \end{aligned}
\end{equation}
We use this functional form in the main manuscript for our analysis. 

\section*{Author Contributions}
R.V.M. contributed to Conceptualization, Writing – original draft, Writing – review \& editing, Validation, Visualization, Data curation, Investigation, Methodology, Software, and Formal Analysis. G.H.M. contributed to Conceptualization, Writing – review \& editing, Methodology, Funding acquisition, Project administration, Resources, and Supervision. R.P. and E.P. contributed to Writing – review \& editing, and Resources. 

\section*{Conflicts of interest}
There are no conflicts to declare.

\section*{Acknowledgements}
The authors would like to thank Lubrizol Inc. for funding and providing the polymer fluids used in this study. 





\bibliography{rsc} 

\end{document}